\documentclass[12pt,onecolumn]{IEEEtran}
\usepackage{amsmath}
\usepackage{amsfonts}
\usepackage{amssymb}
\usepackage{dsfont}
\usepackage{enumerate}
\usepackage{multirow}
\usepackage{color}
\usepackage{graphicx}
\usepackage{epstopdf}
\usepackage{caption}
\usepackage{cases}
\usepackage{subfigure}
\usepackage{cite}

\newtheorem{lemma}{Lemma}
\newtheorem{theorem}{Theorem}

\newtheorem{proposition}{Proposition}
\allowdisplaybreaks[4]
\begin{document}
\baselineskip 4.0ex

\title{Base Station Cooperation in Millimeter Wave Cellular Networks: Performance Enhancement of Cell-Edge Users
\thanks{$^\dagger$ The authors are with the
School of Electronics and Information Engineering, and also with the Ministry
of Education Key Lab for Intelligent Networks and Network Security, Xi'an Jiaotong
University, Xi'an, 710049, Shaanxi, P. R. China.
Email: {\tt  xjbswhm@gmail.com, XJTU-HuangKW@outlook.com}.
}
\thanks{$^\ddagger$ The author is with the School of Electrical and Information Engineering,
Jinan University (Zhuhai Campus), Zhuhai 519070, P. R. China.
Email: {\tt
theodoros.tsiftsis@gmail.com}.
}
}
\author{Hui-Ming Wang$^\dagger$, \emph{Senior Member, IEEE}, Ke-Wen Huang$^\dagger$, and
Theodoros A. Tsiftsis$^\ddagger$, \emph{Senior Member, IEEE}
}
\maketitle

\begin{abstract}
    Millimeter wave (mmWave) signals are much more sensitive to blockage, which results in a significant increase of the outage probability, especially for the users at the edge of the cells.
    In this paper, we exploit the technique of base station (BS) cooperation to improve the performance of the cell-edge users in the downlink transmission of mmWave cellular networks.
    We design two cooperative schemes, which are referred to as fixed-number BS cooperation (FNC) scheme and fixed-region BS cooperation (FRC) scheme, respectively.
    In FNC scheme, the cooperative BSs consist of the $M$ nearest BSs around the served cell-edge users, and in FRC scheme, the cooperative BSs include all the BSs located within a given region.
    We derive the expressions for the average rate and outage probability of a typical cell-edge user located at the origin based on the stochastic geometry framework.
    To reduce the computational complexity of our analytical results for the outage probability, we further propose a Gamma approximation based method to provide approximations with satisfying accuracy.
    Our analytical results incorporate the critical characteristics of mmWave channels, i.e., the blockage effects, the different path loss of LOS and NLOS links and the highly directional antenna arrays.
    Simulation results show that the performance of the cell-edge users is greatly improved when mmWave networks are combined with the technique of BS cooperation.
\end{abstract}
\IEEEdisplaynontitleabstractindextext
\IEEEpeerreviewmaketitle
\section{Introduction}
Recent years has witnessed the dramatically increasing demands on higher wireless data rate.
The conventional wireless networks are incapable of  supporting such an explosive growth in wireless data traffics because of the limited microwave spectrum (below $6$ GHz).
Therefore, due to the wide available bandwidth in the millimeter wave (mmWave) spectrum (up to $300$GHz), mmWave communication has been regarded as one of the most promising technologies for realizing future mobile communication networks \cite{What5GBe,A.Pi2011,FiveDisruptive,T.Pappaport2013,S.Rangan2014},

According to recent field measurements \cite{T.Pappaport2013},
many new characteristics emerge in mmWave bands compared to the conventional microwave bands.
One of the major features lies in the significant path loss due to the higher frequencies.
To compensate the higher path loss, highly directional beamforming antenna arrays are generally  equipped at the base stations (BSs).
%The highly directional beam overcomes the serious attenuation of the intended signal
%and keeps the interfering power under a low level. Therefore, mmWave networks experience less interference than the conventional microwave networks \cite{S.Rangan2014}.
Another important feature  is that  mmWave signal is more sensitive to the blockage effects due to the poor diffraction. According to the field measurements, path loss laws of line-of-sight (LOS) and non-line-of-sight (NLOS) links are significantly different \cite{T.Pappaport2013,M.R.Akdeniz2014}.

In view of these new characteristics in mmWave bands, many works were focused on evaluating the performance of the mmWave networks to provide a guideline for BS deployment design.
In \cite{T.Pappaport2013,M.R.Akdeniz2014},
spatial statistical models of the mmWave channel are constructed based on the measured data. Under these models, the signal-to-interference-and-noise ratio (SINR) and the transmission rate of the mmWave systems have been evaluated in \cite{M.R.Akdeniz2014}.
The network-wide performance of mmWave cellular networks was also theoretically investigated, e.g., coverage probability and average rate in \cite{T.Bai2015}, and the mean interfering power and SINR in \cite{V.Petrov2017}, where a stochastic geometry framework has been adopted to model the topological randomness of the future cellular networks due to the
increasingly dense and irregular BS deployments \cite{FiveDisruptive}.
The works mentioned above have shown that mmWave networks have a great potential to support the enormous increase of wireless data traffics. However, their results are on  the average performance of general users over the whole network. As pointed out in \cite{T.Pappaport2013,S.Rangan2014}, mmWave signals are severely vulnerable to shadowing, which results in high probability of signal outages at the \emph{cell-edge users}. Therefore, the performance of the cell-edge users is the major bottleneck, which becomes a great concern of mmWave cellular networks.

To protect the mmWave signal transmissions from shadowing and ensure reliable communication links for the cell-edge users, a promising solution is the BS cooperation \cite{M.Sawahashi2010,D.Lee2012,X.Tao2012}.
With BS cooperation, each cell-edge user can connect to multiple BSs, and even if the signal strength from one of its serving BSs is poor, the other cooperative BSs may still collaboratively provide a sufficiently high signal power level, e.g., there may be not only one intended links in LOS state.
Besides, the BS cooperation transforms the potential strong interferers, e.g., interfering links in LOS state, to cooperative signal sources, which significantly suppresses the interfering power.
Motivated by this observation, in this paper, we investigate the performance of cell-edge users in BS cooperation-aided mmWave cellular networks based on the stochastic geometry framework.

\subsection{Related Works}
To analyze the performance of the mmWave networks, theoretical channel models have been proposed to describe the new propagation characteristics in mmWave bands \cite{T.Bai2014,T.Bai2015,M.Renzo2015,S.Singh2015,A.Thornburg2016,E.Turgut2017TCOM}.
In \cite{T.Bai2014,T.Bai2015}, the authors proposed a LOS ball model to approximate the irregular LOS region, which was shown to be flexible yet accurate enough to capture the features of the blockage effects in mmWave bands by the field measurements \cite{S.Singh2015}.
The LOS ball model was further extended to the two-ball-based blockage model in \cite{M.Renzo2015} and the multiple-ball-based blockage model in \cite{E.Turgut2017TCOM} to account for the three different states of each link, i.e.,  LOS, NLOS and outage.
Based on the established theoretical channel models, the network-wide system performance of the mmWave networks was investigated, e.g., in \cite{S.Singh2015,A.Thornburg2016,C.Wang2016TWC}.

The stochastic geometry framework has been widely  adopted to analyze  the network-wide performance of cellular networks, including the works in \cite{M.Haenggi2009,T.-X.Zheng2017TCOM,J.Andrews2011,P.Madhusudhanan2012,P.Madhusudhanan2016}.
In \cite{M.Haenggi2009,J.Andrews2011}, the authors modeled the random locations of the BSs as a homogeneous poisson point process (HPPP). And based on this model, the outage probability of a typical user was derived in single tier cellular network \cite{J.Andrews2011} and in multiple tiers heterogeneous networks (HetNets) \cite{P.Madhusudhanan2012,T.-X.Zheng2017TCOM,P.Madhusudhanan2016}.
Compared with the traditional grid model, using HPPP to model the locations of BSs and mobile users provides analytical tractability while guarantees satisfying accuracy of the analytical results \cite{M.Haenggi2009}.

%Device-centric architecture is one of the developmental tendency in future wireless networks, wherein BS cooperation plays an vitally important role \cite{FiveDisruptive}.
%Though many different kinds of cooperation schemes coexist in recent works, for example, coordinated scheduling and coordinated beamforming (CS/CB) \cite{H.Dahrouj2010}, joint transmission (JT) \cite{M.Sawahashi2010} and transmission point selection (TPS) \cite{D.Lee2012}, the main principle behind them are the similar, which can be interpreted as the joint processing of the intended signal and the joint management of the interferences.
%Both analytical results and field trials have show that BS cooperation is an efficient technique to suppress the interferences and improve the spectral efficiency \cite{X. Tao2012}, \cite{R.Irmer2011}.

Based on the stochastic geometry model, BS cooperation in conventional microwave networks has already been extensively studied in the past few years
\cite{G.Nigam2014,G.Nigam2015,G.Nigam2015GLOBECOM,R.Tanbourgi2014_1,
R.Tanbourgi2014_2,W.Nie2016,Q.Cui2018TCOM,X.Yu2017WCL}.

In \cite{G.Nigam2014}, BS cooperative downlink transmissions in HetNets were investigated.
The authors of \cite{G.Nigam2014} derived the coverage probabilities and the diversity gains for the general users in multi-tier networks and the worst-case users in single-tier networks.
In \cite{G.Nigam2015} and \cite{G.Nigam2015GLOBECOM}, the performance of the general and worst-case users in spatiotemporal BS cooperation-aided stochastic networks were investigated, respectively. It was shown that the spatiotemporal BS cooperation is an effective way to increase the coverage and suppress the interference.
The authors of \cite{Q.Cui2018TCOM} considered the joint transmission (JT), dynamic point selection/dynamic point blanking (DPS/DPB), and the combination of JT and DPB in BS cooperation-aided cellular networks, and investigated the meta distribution of the signal-to-interference ratio.
The authors of \cite{X.Yu2017WCL} investigated the success probabilities in downlink HetNets with coherent joint transmission (CJT), and upper bounds on the success probabilities of the general and worst-case users  were derived.
In \cite{R.Tanbourgi2014_1} and \cite{R.Tanbourgi2014_2}, a tractable model for analyzing the performance of non-coherent joint transmission (NCJT) was proposed, and the distribution of SINR was derived. The authors of \cite{W.Nie2016} derived the spectral efficiency under the NCJT BS cooperation in general HetNets, and optimized the received signal strength thresholds of each tier to obtain a higher spectral efficiency.

\subsection{Motivations and Contributions}
Though the works mentioned above illustrated the advantages of the BS cooperation, the conclusions only fit with the conventional microwave networks. The new channel characteristics in mmWave makes the performance of the BS cooperation in  mmWave wireless networks need to be re-evaluated.
We have to point out that compared with the microwave networks, BS cooperation is much more important for the mmWave networks.
This is because the link quality in mmWave band can be significantly reduced due to the blockage effect which is not observed in microwave band. The blockage effect can be compensated by letting more than one BS to serve the users, which increases the probability of having reliable LOS links.
Recently, the authors of \cite{D.Maamari2016} applied the BS cooperation to the mmWave wireless networks to improve the network-wide coverage probability.
However, the results in  \cite{D.Maamari2016} suit to general users but not typically to the cell-edge users, the performance of whom poses the major bottleneck of the mmWave cellular networks.
Motivated by these observations and based on the channel models for the mmWave bands established in the previous works, the performance of the cell-edge users in BS cooperation-aided mmWave cellular networks is analytically studied  under the stochastic geometry framework.
The main contributions of this work are summarized as follows:
\begin{enumerate}
\item We improve the performance of the cell-edge users in mmWave cellular networks by using the technique of BS cooperation.
    Two different cooperation schemes are proposed to serve the cell-edge users, i.e., \emph{fixed-number cooperation (FNC)} and \emph{fixed-region cooperation (FRC)}. Note the the proposed geometry-based BS association strategies should be more practical than the instantaneous received signal strength (IRSS) based association strategy in \cite{D.Maamari2016}. In general, estimating the IRSS from multiple BSs causes much communication overhead, which is especially much more inefficient in mmWave band, because the mmWave links usually have smaller channel coherent time than that in microwave links.
\item  For both strategies, the expressions for average rate and outage probability are derived, while in \cite{D.Maamari2016}, only the coverage probability is investigated.
    We show that the performance of the cell-edge users has a significant promotion with the help  of BS cooperation.
    However, we show that calculating the analytical results of the outage probability is complicated and time-consuming, which motivates us to find a more efficient way to evaluate the outage probability.
\item  To efficiently evaluate the outage probability, we resort to the technique of Gamma approximation \cite{R.W.Heath2013} to obtain an approximation of the outage probability, which significantly differentiate our work from  \cite{D.Maamari2016}. Besides, different from the general idea of Gamma approximation, where the signal power is approximated by only one Gamma random variable (RV), we approximate the received signal power by using two different Gamma RVs to separately characterize the statistics of LOS and NLOS links.
    Numerical results show that the approximate results are of high accuracy and computational efficiency.
\end{enumerate}
{
We note that the proposed cell-edge user model in this paper is different from the
worst-case user model proposed in \cite{G.Nigam2014}. In the latter, the
users that are located at the Voronoi vertices formed by the BSs are referred to as
the worst-case users.
However, when the locations of several BSs are close to each other (which is
possible from the perspective of HPPP stochastic network), the worst-case users at the Voronoi vertices formed by these BSs are hence possible to be very close to these BSs.
The proposed cell-edge user model eliminates
this case by setting an exclusive region of the BSs around the users. The details of the cell-edge model will be presented in the next section.
}

\subsection{Paper Organization and Notations}
In Section II, mmWave channel characteristics and some basic assumptions in this paper are detailedly introduced.
In Section III, the BS cooperation schemes and the performance metrics are presented.
In Section IV and V, the average rate and outage probability of a typical cell-edge user in BS cooperation-aided mmWave cellular networks are derived, respectively. In section VI, numerical examples are presented. Finally, in Section VII, we conclude the paper.

\emph{Notations}: $\left|\cdot\right|$ denotes the cardinality of a set.
$\Gamma\left(s\right)$ denotes the Gamma function.
${}_2F_1\left(a,b;c;z\right)$ denotes the Gauss hypergeometric function \cite[Eq. 9.100]{Book:ISG2007}.
The factorial of a non-negative integer $M$ is denoted by $M!$.
$h\sim \mathrm{Gamma}\left(k,m\right)$ denotes Gamma-distributed RV with shape parameter $k$ and scale parameter $m$.
$\mathcal{L}_{X}\left(z\right)$ is the Laplace transform (LT) of a RV $X$.
$\mathbb{E}\left(\cdot\right)$ and $\mathbb{D}\left(\cdot\right)$ denote the expectation and variance operation. $\mathcal{B}\left(x,d\right)$ denotes the ball region whose center is $x$ and radius is $d$.
$\mathcal{R}\left(x,D_O,D_I\right)$ denotes the annular region whose center is $x$ and the outer and inner radius are  $D_O$ and $D_{I}$, respectively.
$\mathrm{Im}\left(\cdot\right)$ denotes the imaginary part of a complex number. For a point process $\Phi$, {\color{blue}$\Phi\left(\mathcal{A}\right)$ denotes the set of the points in $\Phi$ that are located within an arbitrary area $\mathcal{A}$, and
$\left|\Phi\left(\mathcal{A}\right)\right|$ denotes the number of the points in $\Phi\left(\mathcal{A}\right)$}.

\section{System Model}
\label{Section2}
In this section, we introduce the system model to evaluate the performance of the cell-edge users in BS cooperation-aided mmWave cellular networks.
We consider the downlink transmissions and
some basic assumptions are provided in the following subsections.

\subsection{BS Layout and User Distribution}
Following \cite{J.Andrews2011}, the BSs are assumed to be spatially distributed in a two-dimensional plane according to a HPPP $\Phi_B$ with density $\lambda_B$.
Using HPPP to model the irregular
BSs locations provides a tractable
approach for characterizing the downlink performance of the
cellular network.
In this paper, we also use the concept of the \emph{average cell radius}, denoted by $\rho\triangleq\sqrt{1/\pi\lambda_B}$, to equivalently represent the density of the BSs.
For analytical tractability, we assume that all the BSs transmit with the same power denoted by $P_T$ in each time/frequency resource block.

For the users, we assume they are distributed as a HPPP with density $\lambda_U$.
The users will connect to their surrounding BSs, and if one BS are connected by multiple users, we
assume the BS will schedule the users to different orthogonal time/frequency resource blocks so
that it only serves one user at a certain time/frequency slot \footnote{
In general, with multiple antennas equipped at the BSs, it is possible that each BS can serve multiple users in the same time/frequency resource block by using the space division multiple access. However, in this case, the interference analysis becomes mathematically intractable, because the interfering beam pattern of an interfering BS is now correlated to the number and the locations of the multiple users.
In this paper, we assume that each BS only serves one user at a certain time/frequency slot for simplicity. Similar assumption is also adopted in existing works such as \cite{J.Andrews2011,T.-X.Zheng2017TCOM,W.Nie2016,D.Maamari2016}.
%{\color{blue}
%As we assumed in Section II-A that each BS only serves one user in a certain time/frequency resource block,
%we have that each cooperative BS only serves the typical cell-edge user at the consider time/frequency slot.
%Note that the assumption of having multiple BSs to serve only one user may not be practical in fully loaded networks where there are a large amount of users due to the lack of time/frequency resources.
%However, this assumption may suits to the following two cases: 1) when the considered cellular network is slightly loaded and there are available time/frequency resources which can be scheduled by the BSs for different users; and 2) when the density of the users is relatively small and each user only competes with a small number of users for the communication resources of a BS.
%Similar assumption is also adopted in \cite{W.Nie2016}.
%}
}.

\subsection{Cell-Edge User Model}
Note that in this paper, we focus on the performance of the cell-edge users.
The cell-edge users are modeled as those users who are sufficiently far away from any BS.
More specifically, a user, located at $x\in \mathcal{R}^2$, is viewed as an edge user if there is no BS located within a distance of $D_E$ from $x$, i.e., $\left|\Phi_B\left(\mathcal{B}\left(x,D_E\right)\right)\right|=0$
\footnote{The proposed cell-edge user model is inspired by the model proposed in \cite{N.Deng2014}.
In \cite{N.Deng2014}, the authors considered a two-tier HetNet, wherein the macro BSs (MBSs) follow a HPPP and pico BSs (PBSs) follow a PHP. A MBS serves the users located within a circle area with a pre-designed  radius around it. PBSs are only deployed outside the coverage regions of MBSs to improve the performance of the users in the coverage holes of the MBSs, or in an other word, the cell-edge users.
}.
To properly characterize the cell-edge users, $D_E$ is defined to be proportional to the average cell radius, i.e., $D_E\triangleq\chi\rho$, where $\chi$ is the scaling factor.
When we set $\chi=1$, an intuitive interpretation for the proposed edge user model is as follows:
any one of the users in the network is expected to be covered by, on average, one BS within the distance of $D_E$, and therefore, for a given user, if therefore is no BS located within $D_E$, the user will be referred to as an cell-edge user.
According to the proposed edge user model above, the locations of the cell-edge users follow a poisson hole process (PHP) \cite{C.h.Lee2012,Z.Yazdanshenasan2016NovTWC}.
To analyze the performance of the cell-edge users, we assume there is a typical cell-edge user located at the origin $o$, which leads to an exclusive ball region, i.e., $\mathcal{B}\left(o,D_E\right)$, for the BSs \footnote{
Though we introduce a scaling factor $\chi$ when determining the cell-edge users, the derived theoretical results apply to an arbitrary value of $\chi$.
%In the proposed cell-edge user model, we introduce a scaling factor $\chi$ when determining the cell-edge users.
%Here, with the scaling factor $\chi$, we not only seizes the basic property of the cell-edge users (e.g. far away from the BSs), but also provides an extra degree of design freedom. More specifically, the value of $\chi$ can be designed to achieve the trade-off between the whole system resources consumption, e.g., the energy efficiency and backhaul capacity, and the proportion of the users that are classified as the cell-edge users who should be served by the BS cooperation, which is, however, out the scope of this paper.
%We also note that the theoretical results for the average rate and outage probability derived in this paper apply to an arbitrary value of $\chi$.
}.

\subsection{Single-Cluster MmWave Channel Model and Directional Beamforming}
\label{DirectionalBeam}
According to the field measurements in \cite{M.R.Akdeniz2014}, the propagation of the mmWave signal is usually in clusters.
The clustered mmWave channel model has been used to analyze and optimize the performance of the mmWave communication systems, e.g., in \cite{M.R.Akdeniz2014,O.E.Ayach2014TWC}.
However, for analytical tractability, following the works in \cite{E.Turgut2017TCOM,C.Wang2016TWC,D.Maamari2016}, we consider the single-cluster channel model in this paper, i.e., we only consider the direct transmission path from the BSs to the users.

To overcome the significant path loss and guarantee sufficient link margins, all the BSs are equipped with highly directional beamforming antenna array. In this paper, the gain pattern of the antenna array $G_B\left(\theta\right)$ is described by the following sectored model,
\begin{align}\label{GainPattern}
G_B\left( \theta \right)=
\begin{cases}
  G_M, & \left| \theta \right|\leq \theta_T,\\
  G_S, & \left| \theta \right|> \theta_T,
\end{cases}
\end{align}
where $\theta$ is the angle of direction, $\theta_T$ is the beam width of the main lobe, and $G_M$ and $G_S$ are the directional gains in the main lobe and side lobe, respectively \footnote{A widely used antenna array is the uniform linear array (ULA) with each antenna element separated half of a wavelength. In this case, the gain pattern  is written as
$G_{ULA}\left( \theta \right) = \frac{1}{K} \sin\left(\frac{K\pi}{2}\cos\left(\theta\right)\right)^2/
\sin\left(\frac{\pi}{2}\cos\left(\theta\right)\right)^2$,
where $K$ is the number of the antennas and $\theta$ is the angle of departure.
We can use \eqref{GainPattern} to approximate the gain pattern of the ULA
by setting $G_M=K$, $G_S=\max_{\frac{2}{K}<x<\frac{4}{K}}~
\sin\left(\frac{K\pi}{2}x\right)^2/
\sin\left(\frac{\pi}{2}x\right)^2$, and
$\theta_T = 4\left(\frac{\pi}{2}-\arccos\left(\frac{2}{K}\right)\right)$.}.
We note that the in practice, the actual beam pattern is generally very complicated and is highly related to the issue of beamforming and procoding design, e.g., see \cite{S.Han2015CM,O.E.Ayach2014TWC} and references therein. In this paper, we mainly focus on evaluating the network-wide performance and therefore, the complicated beam pattern is approximated by the sectored model in \eqref{GainPattern} for mathematical simplicity as in \cite{T.Bai2015,M.Renzo2015,S.Singh2015,E.Turgut2017TCOM,C.Wang2016TWC}.
Assume that each BS can always orient its main lobe towards the intended receiver, and thus the directional gains of the intended links are always $G_M$
\footnote{
To enable the directional transmission, the BSs should first obtain the directions of their intended users. This can be realized by an aforehand beam training procedure, e.g., in \cite{C.Jeong2015CM,Y.Li2017TWC}. For simplicity, in this paper, we assume the BSs have already obtained the directions of their intended users.
}.
For each interfering link, due to the HPPP distribution, the angle $\theta$ with respect to (w.r.t.) the typical user is independently and uniformly distributed in $[-\pi,\pi]$, which means that the direction gain $G_B\left( \theta_j \right)$ from the $j^{\mathrm{th}}$ interfering link becomes a binary RV with probability mass function (PMF) given as
\begin{align}\label{GainPatternB}
G_B\left( \theta_j \right)=
\begin{cases}
  G_M, & \mathbb{P}\left\{ G_B\left( \theta_j \right)=G_M \right\}=\frac{\theta_T}{2\pi},\\
  G_S, & \mathbb{P}\left\{ G_B\left( \theta_j \right)=G_S \right\}=\frac{2\pi-\theta_T}{2\pi}.
\end{cases}
\end{align}

\subsection{Blockage Model}
In this paper, we adopt the LOS ball blockage model proposed in \cite{T.Bai2015}.
Define $q_L\left( d \right)$ as the probability that an arbitrary link of distance $d$ is LOS, then
\begin{align}\label{BlockageModel}
q_L\left( d \right)=
\begin{cases}
  C, & d \leq D, \\
  0, & d > D,
\end{cases}
\end{align}
where $D$ is the radius of the approximate LOS region, and $0\leq C\leq 1$ can be physically interpreted as the average proportion of LOS links in the LOS region. Accordingly, the probability that an arbitrary link of distance $d$ is NLOS is defined as $q_N\left( d \right)= 1- q_L\left( d \right)$.
For the ease of analysis, we assume that that states (LOS or NLOS) of different wireless links are independent.
It was shown that this model is simple yet flexible enough to describe the  impacts of the blockage effects in mmWave bands \cite{T.Bai2015}.

\subsection{Path Loss Model}
It has been pointed out in \cite{M.R.Akdeniz2014} that the path loss laws are significantly different between LOS and NLOS links in mmWave bands. For a link of distance $d$, the path loss can be expressed as
\begin{align}\label{PathLossModel}
L\left( d \right)=
\begin{cases}
  C_Ld^{-\alpha_L}, & \textrm{LOS link}, \\
  C_Nd^{-\alpha_N}, & \textrm{NLOS link},
\end{cases}
\end{align}
where $\{\alpha_L, C_L\}$ and $\{\alpha_N, C_N\}$ are the path loss exponents and the path loss gain at unit distance of LOS and NLOS links, respectively \footnote{Typical value of $\{\alpha_L,C_L\}$ and $\{\alpha_N,C_N\}$ can be found in
\cite[\textrm{Table I}]{M. R. Akdeniz2014}, and, usually, they satisfy that $2\leq\alpha_L<\alpha_N$ and $C_L>C_N$.}.

\subsection{Small-Scale Fading}
Field experiments show that the distributions of the small-scale fading in LOS and NLOS links are also different due to the different propagation environments \cite{M. R. Akdeniz2014,T.Bai2015}.
Using $h$ to denote the small-scale power fading, we have
\begin{align}\label{SmallScaleFadingModel}
h=
\begin{cases}
  h^{(L)}, & \textrm{LOS link}, \\
  h^{(N)}, & \textrm{NLOS link},
\end{cases}
\end{align}
where $h^{(L)}$ and $h^{(N)}$ are small-scale power fading factors in LOS and NLOS links, respectively.
Nakagami fading is assumed throughout this paper because both of the rich or non-rich scattering environments can be modeled by changing the parameters of the Nakagami distribution.
Based on this assumption, the small-scale power fading factors becomes Gamma RVs, i.e., $h^{(\nu)}\sim \mathrm{Gamma}\left(N_{\nu},\frac{1}{N_{\nu}}\right)$ for $\nu \in \{L,N\}$, where $N_L$ and $N_N$ are the parameters for LOS and NLOS links, respectively.

\section{BS Cooperation Strategy and Performance Metrics}
\label{Section:CooperativeStrategy}
Define $\Omega$ as the set of cooperative BSs serving the typical cell-edge user.
We consider two kinds of cooperation schemes, i.e., the FNC and FRC schemes. We present the details about the two cooperative schemes in the following.

\subsection{FNC and FRC Schemes}
\begin{figure}[tp]
\begin{center}
\includegraphics[width=5.5 in]{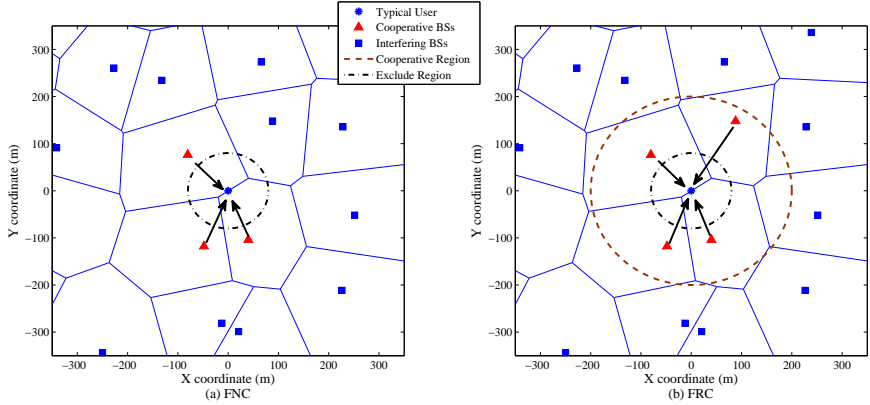}
\caption{(a) FNC scheme with $M=3$. the nearest $3$ BSs cooperatively serve the typical user; (b) FRC scheme. All the BSs  within the cooperative region collaboratively serve the typical user.}\label{Fig:model}
\end{center}
\end{figure}

In the FNC scheme, we assume that $\Omega$ consists of the $M$ nearest BSs. In FRC, $\Phi_C$ consists of all the BSs located within a cooperative region denoted as ball region $\mathcal{B}\left(o,D_{\mathrm{co}}\right)$. Due to the exclusive ball region $\mathcal{B}\left(o,D_E\right)$, the  cooperative region of FRC is an annular region, i.e., $\mathcal{R}\left(o,D_E,D_{\mathrm{co}}\right)$.
Fig. \ref{Fig:model} illustrates these two cooperative strategies.
In fact, the significant path loss and the severe blockage effects in mmWave bands cause that the reliable links usually only exist when the distances between the transceivers are short.
Therefore, both FNC and FRC attempt to associate the users to the BSs that are not far away from them.
The basic difference between the FNC and FRC schemes lies in the fact that the FNC scheme always associates the typical cell-edge user with $M$ surrounding BSs regardless of their distances to the cell-edge user, while
in the FRC scheme, the cooperative BSs are restricted to be sufficiently close to the typical cell-edge user which results in the randomness of the number of the cooperative BSs in the FRC scheme.
The advantage of the FNC scheme is that the typical cell-edge user is always guaranteed to connect to the network (though perhaps with a poor link quality), while the FRC scheme may lead to a disconnecting state at the typical cell-edge user because it is possible that there is no BS located within the cooperative region.
However, it is also possible that the average performance of the FRC scheme outperforms that of the FNC scheme.
This is because that the random number of the cooperative BSs in the FRC shceme may possibly exceed that in the FNC scheme, which improves the performance of the typical cell-edge user.
Note that we only focus on the performance analysis of the typical cell-edge user located at the origin, all our analytical results suit to the cases when there is no BS within the distance of $D_E$ around the origin.
Conditioning on the fact that there is no BS in $\mathcal{B}\left(o,D_E\right)$,
we have the following lemmas, which will be used in the derivations of the main conclusions.

\begin{lemma}
The number of cooperative BSs in FRC scheme is a RV with PMF given by
\begin{align}
\label{CooperativeNumFRC}
\mathbb{P}\left\{ \left|\Omega\right| =n\right\}= \left(\lambda_B \pi \left(D_{\mathrm{co}}^2-D_E^2\right)\right)^ne^{-\lambda_B \pi \left(D_{\mathrm{co}}^2-D_E^2\right)}\big/{n!}.
\end{align}
\begin{IEEEproof}
Directly obtained from the property of the HPPP.
\end{IEEEproof}

\end{lemma}

\begin{lemma}
\label{M+1NearstBSPDF}
Denote $D_{i}$ as the distance between the typical cell-edge user and the $i^{\mathrm{th}}$ nearest BS. Conditioning on there is no BS located within $\mathcal{B}\left(o,D_E\right)$, the probability density function (PDF) of $D_{M+1}$ is given by
\begin{align}\label{M+1PDF}
  f_{D_{M+1}}\left( r \right) = \mathbb{I}\left(r>D_E\right)2\lambda_B \pi r\left( \lambda_B\pi \left(r^2-D_E^2\right) \right)^M
   \exp \left( -\lambda_B\pi \left(r^2-D_E^2\right)\right)\big/{M!}
  .
\end{align}
where $\mathbb{I}\left(\cdot\right)$ denotes the indicator function.
\end{lemma}
\begin{IEEEproof}
The conditional cumulative distribution function (CDF) of $D_{M+1}$ can be calculated as
\begin{align}
\label{M+1CDF}
&\mathbb{P}\left\{D_{M+1}\leq r|\left|\Phi_B\left(\mathcal{B}\left(o,D_E\right)\right)\right|=0\right\}=\mathbb{P}\left\{ \left|\Phi_B \left( \mathcal{R} \left(o,D_E,r\right)\right)\right|> M \right\} \nonumber \\
&=1-\exp \left( -\lambda _B \pi \left(r^2-D_E^2\right) \right)\sum\nolimits_{k=0}^{M}\left( \lambda _B\pi \left(r^2-D_E^2\right) \right)^k\big/{k!},
\end{align}
where $\Phi_B$ is the point process of the BS defined in Section II-A.
Then the conditional PDF can be obtained by taking a derivative w.r.t. $r$.
\end{IEEEproof}

\begin{lemma}
\label{Lemma:LocationOfCooperativeBSFNC}
In FNC scheme, conditioning on  $D_{M+1}=r$, there are $M$ BSs in the region $\mathcal{R}\left(D_E,r\right)$.
We refer to these $M$ cooperative BSs as $\mathds{BS}_1$, $\mathds{BS}_2$, $\cdots$, $\mathds{BS}_M$, which is independent of their distances to the typical cell-edge user.
Denote the distances between $\mathds{BS}_1$, $\mathds{BS}_2$, $\cdots$, $\mathds{BS}_M$ and the typical cell-edge user as $D^{(1)},D^{(2)},\cdots,D^{(M)}$, respectively.
Then, we have $\left\{D^{(k)}\right\}_{k=1}^{M}$ being identical and independent distributed (i.i.d.) RVs with conditional PDFs given by
%\footnote{The relation between $\{D^{(k)}\}_{k=1}^M$ and $\{D_{k}\}_{k=1}^M$ can be written as $D_{k} = \max_{ d\in  \{D^{(i)}\}_{k=1}^M \setminus \{D_1,D_2,\cdots,D_{k-1}\}  } d,~~k = 1,2,\cdots,M$, i.e., $\{D_{1},D_{2},\cdots,D_{M}\}$ is an ordered version of $\{D^{(1)},D^{(2)},\cdots,D^{(M)}\}$.}
\begin{align}\label{DistanceConditionPDF}
f_{D^{(k)}|D_{M+1}=r}\left(x\right) = \mathbb{I}\left(D_E<x\leq r\right)2x\big/\left(r^2-D_E^2\right), \quad \forall k = 1,2,\cdots,M.
\end{align}
\end{lemma}
\begin{IEEEproof}
According to the relationship between the HPPP and the uniform binomial point process (UBPP) \cite[Chapter 2.3.1]{Book:SG2013}, conditioning on the fact that there are $M$ BSs in the region $\mathcal{R}\left(D_E,r\right)$, the conditional HPPP in $\mathcal{R}\left(D_E,r\right)$ is equivalent to an UBPP, i.e.,
the $M$ BSs are independently uniformly distributed in $\mathcal{R}\left(D_E,r\right)$.
Therefore, we have $\mathbb{P}\left\{D^{(k)} \leq x\right\}=\frac{x^2-D_E^2}{r^2-D_E^2}$. The conditional PDFs $f_{D^{(k)}|D_{M+1}=r}\left(x\right)$ can be derived in \eqref{DistanceConditionPDF} by taking a derivative w.r.t. $r$.
\end{IEEEproof}

\subsection{Performance Metrics}
In this paper, we adopt the non-coherent joint transmission (NCJT) scheme, where the cooperative BSs transmit the same data to the typical user non-coherently \cite{R.Tanbourgi2014_1,R.Tanbourgi2014_2,W.Nie2016}. This scheme enjoys less requirement on strict synchronization and accurate channel state information (CSI) compared with the coherent joint transmission scheme (CJT) \cite{R.Tanbourgi2014_1}, and therefore is more practical.
Following the assumption of NCJT in \cite{R.Tanbourgi2014_1}, the signal power and interfering power received at the typical cell-edge user are respectively given by
\begin{subequations}
\label{SignalInterference}
\begin{align}
  \label{SignalPower}
  T &\triangleq \sum_{i\in\Omega} P_T G_M h_i L_i\left( D_i \right),\\
  I &\triangleq \sum_{j\in\Phi_B\setminus\Omega} P_T G_B\left(\theta_j\right) h_j L_j\left( D_j \right),
  \label{InterPower}
\end{align}
\end{subequations}
where $L_i\left( D_i \right)$, $h_i$ and $D_i$ for $\forall i \in \Phi_B$ are the path loss, small-scale power fading and the distance between the $i^{\mathrm{th}}$ nearest BS and the typical cell-edge user at the origin $o$, respectively, $\theta_j$ and $G_B\left(\theta_j\right)$ are the angle and the directional gain of the interference from the $j^{\mathrm{th}}$ interfering BS, respectively.
%\footnote{Note that in practice, the interference is only caused by the transmissions in
%the same time/frequency resource block, while in \eqref{InterPower}, we assume that all the interfering BSs are active and will cause interference in the time/frequency resource block used by the typical cell-edge user. This can be viewed as the worst case for the typical cell-edge user. Similar assumption is implicitly adopted in \cite{G.Nigam2014,X.Yu2017WCL,D.Maamari2016}.}.
According to Section \ref{DirectionalBeam},
$G_B\left(\theta_j\right)$ is a binary RV with PMF given as \eqref{GainPatternB}.
Based on the assumptions made above, the SINR of the typical user is given by
$\mathrm{SINR} = \frac{T}{I + N_0}$,
where $N_0$ is the power of thermal noise and we have $N_0 (dB) = -174 + 10\mathrm{log}_{10}\left(\mathrm{BW}\right)+ F$ with $\mathrm{BW}$ and $F$ denoting the bandwidth and the noise figure, respectively.
{\color{blue}Note that in practice, only the BSs that are active in the same frequency/time resource block will cause interference. However, to exactly characterize the point process of the interfering BSs is very diffcult, and the distribution of the the interfering BSs  is highly related to the density of the users and the resource allocation policy of the network. Therefore, following existing works \cite{G.Nigam2014,X.Yu2017WCL,D.Maamari2016}, each BS in $\Phi_B\setminus\Omega$ is assumed to be an interferer for analytical tractability, which can be viewed as the worst case for the typical cell-edge user.}
%Nevertheless, the system performance under such worst case will not deviate much from that in actual case  as we will show in Section VI.

In this paper, we evaluate the average rate and the outage probability of the typical cell-edge user in BS cooperation-aided mmWave cellular networks.
The average rate is given by
\begin{align}\label{AverageRate}
\bar{R} = \mathbb{E}\left[\ln\left(1+\mathrm{SINR}\right)\right],
\end{align}
where the expectation operation is taken w.r.t. the random locations of BSs, the LOS or NLOS state of each link, the angle of all the interfering links and the small-scale power fading.
The outage probability is defined as the probability that the $\mathrm{SINR}$ is smaller than a threshold $\tau$, i.e.,
\begin{align}\label{OutageProbability}
\mathcal{P}_{O}\left(\tau\right) = \mathbb{P}\left\{\mathrm{SINR} \leq \tau \right\}.
\end{align}

For the sake of simplicity and clarity, in the following of this paper, we use $x^{M,\mathrm{FN}}$ and $x^{D_{CO},\mathrm{FR}}$ for $x\in \{T,I,\mathrm{SINR},\Omega,\bar{R}, \mathcal{P}_{O}\left(\tau\right)\}$ to represent the corresponding physical quantities under the FNC and FRC schemes, respectively.

\section{Average Rate of the Typical Cell Edge User}
\label{Section3}
In this section, the average rates of the typical cell-edge user in BS cooperation-aided mmWave cellular network are derived for both FNC and FRC schemes.
Before presenting the details, we first introduce an important Lemma which will be used in the following.
\begin{lemma}
\label{LemmaRate}
For independent RVs $X$ and $Y$ satisfying $X\geq0$ and $Y>0$, we have
\begin{align}
\mathbb{E}_{X,Y}\left(\ln\left(1 + \frac{X}{Y}\right)\right)=\int_0^{+\infty}\left(1 - \mathcal{L}_X\left(z\right)\right)\mathcal{L}_Y\left(z\right)\frac{\mathrm{d}z}{z}.\label{TheOriRate}
\end{align}
\end{lemma}
\begin{IEEEproof}
According to \cite[Lemma 1]{K. A. Hamdi2008}, we have
\begin{align}
\mathbb{E}_{X,Y}\left(\ln\left(1 + \frac{X}{Y}\right)\right) = \mathbb{E}_{X,Y}\left(\int_0^{+\infty}\left(1 - \exp\left(-\frac{X}{Y}t\right)\right)\exp\left(-t\right)\frac{\mathrm{d}t}{t}\right).
\end{align}
By making a change of variable $\frac{t}{Y}\rightarrow z$ and exchanging the order of integration and expectation, we obtain \eqref{TheOriRate}.
\end{IEEEproof}

With Lemma \ref{LemmaRate}, we now provide the average rates under the two cooperative strategies.
\subsection{Average rate of the typical cell-edge user in FNC scheme}
The following theorem presents the average rate of the typical cell-edge user in FNC scheme.
\begin{theorem}
$\bar{R}^{M,\mathrm{FN}}$ is given by
\begin{align}
\label{AverageRateFNC}
\bar{R}^{M,\mathrm{FN}}& = \int_{D_E}^{+\infty }f_{M+1}\left( r \right) \Bigg\{ \int_{0}^{+\infty}\left( 1- \mathcal{L}_{T^{M,\mathrm{FN}}|r}\left(z\right)\right) \mathcal{L}_{I^{M,\mathrm{FN}}|r}\left(z\right) \exp \left( -N_0z \right) \frac{\mathrm{d}z}{z} \Bigg\}\mathrm{d}r,
\end{align}
where we define $\mathcal{L}_{X|r}\left(z\right) \triangleq \mathbb{E}_{X}\left[ \exp \left( -Xz \right)\left| D_{M+1}=r \right. \right]$ as the conditional LT of a RV $X$.
\end{theorem}
\begin{IEEEproof}
Using \eqref{AverageRate}, we first have
$
\bar{R}^{M,\mathrm{FN}} = \mathbb{E}_{T^{M,\mathrm{FN}},I^{M,\mathrm{FN}}}\left(\ln\left(1 + \frac{T^{M,\mathrm{FN}}}{I^{M,\mathrm{FN}} + N_0}\right)\right).
$
Note that in this equation, $T^{M,\mathrm{FN}}$ and $I^{M,\mathrm{FN}}$ are mutually dependent, therefore, Lemma \ref{LemmaRate} is not directly applicable.
To deal with it, we resort to the law of total probability, and then $\bar{R}^{M,\mathrm{FN}}$ can be rewritten as
\begin{align}
\label{AverageRateFNCProof2}
\bar{R}^{M,\mathrm{FN}} = \int_{D_E}^{+\infty }f_{M+1}\left( r \right)\left\{\mathbb{E}_{T^{M,\mathrm{FN}},I^{M,\mathrm{FN}}}\left(\ln\left(1 + \frac{T^{M,\mathrm{FN}}}{I^{M,\mathrm{FN}} + N_0}\right)\Big|D_{M+1}=r\right)\right\}\mathrm{d}r.
\end{align}
In \eqref{AverageRateFNCProof2}, conditioning on $D_{M+1}=r$, $T^{M,\mathrm{FN}}$ and $I^{M,\mathrm{FN}}$ are mutually independent.
By applying Lemma \ref{LemmaRate} to \eqref{AverageRateFNCProof2}, we obtain \eqref{AverageRateFNC}.
\end{IEEEproof}

The calculation of \eqref{AverageRateFNC} requires the conditional LTs of $T^{M,\mathrm{FN}}$ and $I^{M,\mathrm{FN}}$, i.e.,
$\mathcal{L}_{T^{M,\mathrm{FN}}|r}\left(z\right)$ and $\mathcal{L}_{I^{M,\mathrm{FN}}|r}\left(z\right)$,
which are given in the following two theorems, respectively.
\begin{theorem}
\label{Lemma:LaplaceTransInFNCStep1}
The LTs of $T^{M,\mathrm{FN}}$ conditioning on $D_{M+1} = r$, i.e., $\mathcal{L}_{T^{M,\mathrm{FN}}|r}\left(z\right)$, is given by
\begin{align}
\mathcal{L}_{T^{M,\mathrm{FN}}|r}\left(z\right)&=\left[ \mathcal{L}_{T_k^{M,\mathrm{FN}}|r}\left(z\right) \right]^M, \quad \forall k\in \{1,2,\cdot\cdot\cdot,M\},\label{FNCSignalLaplaceTrans}\\
\mathcal{L}_{T_k^{M,\mathrm{FN}}|r}\left(z\right)& = \frac{ \sum_{\nu\in \{L,N\}}p_\nu\Lambda_{M,\nu}\left(z,\mathrm{min}\left(r,D\right),D_E\right)
+\mathbb{I}\left(r>D\right)\Lambda_{M,N}\left(z,r,D\right) }{r^2-D_E^2},
\label{FNCSignalLaplaceTransTk}
\end{align}
where
$T_k^{M,\mathrm{FN}}\triangleq P_T G_M h_{(k)} L_{(k)}\left(D^{(k)}\right)$ with $L_{(k)}\left(D^{(k)}\right)$, $h_{(k)}$, and $D^{(k)}$ being the path loss, small scale power fading, and distance between $\mathds{BS}_k$ (defined in Lemma \ref{Lemma:LocationOfCooperativeBSFNC}) and the typical cell-edge user, respectively,
$p_L\triangleq C$ and $p_N\triangleq 1-C$ are  the probabilities that a BS within the LOS region is in LOS and NLOS state, respectively, and
\begin{subequations}
\begin{align}
\Lambda_{\mu,\nu}\left(z,x,y\right)&\triangleq\int_{y}^x 2t\mathcal{L}_{h^{(\nu)}}\left(za_{\mu,\nu}t^{-\alpha_{\nu}}\right) \mathrm{d}t\nonumber
=\int_{y}^x \frac{2t}{\left(1+\frac{1}{N_{\nu}}za_{\mu,\nu}t^{-\alpha_{\nu}}\right)^{N_{\nu}}} \mathrm{d}t\\
&=\frac{2}{2+\alpha_\nu N_\nu}\left(\frac{za_{\mu,\nu}}{N_\nu}\right)^{-N_\nu}\left(\Delta_{\mu,\nu}
\left(z,x\right) - \Delta_{\mu,\nu}\left(z,y\right)\right), \label{LambdaFunc}\\
\Delta_{\mu,\nu}\left(z,t\right)&\triangleq t^{\frac{2}{\alpha_\nu}(N_\nu+\frac{2}{\alpha_\nu})}{}_2F_1\left(N_\nu,N_\nu+\frac{2}{\alpha_\nu};N_\nu+\frac{2}{\alpha_\nu}+1;
-\frac{N_\nu t^{\frac{2}{\alpha_\nu}}}{za_{\mu,\nu}}\right),
\end{align}
\end{subequations}
with $a_{\mu,\nu}\triangleq P_TG_{\mu}C_{\nu}$ for $\forall \mu\in\{M,S\}$ and $\forall \nu\in\{L,N\}$.
\end{theorem}
\begin{IEEEproof}
Due to the fact that $\mathds{BS}_1$, $\mathds{BS}_2$, $\cdots$, $\mathds{BS}_M$ are exactly the cooperative BSs in $\Omega^{M,\mathrm{FN}}$, we have $T^{M,\mathrm{FN}} = \sum_{k = 1}^M T_k^{M,\mathrm{FN}}$.
According to Lemma \ref{Lemma:LocationOfCooperativeBSFNC},  under the condition that $D_{M+1}= r$, $\left\{D^{(k)}\right\}_{k=1}^{M}$, are i.i.d. RVs. Therefore, we can obtain \eqref{FNCSignalLaplaceTrans}.
The derivation of \eqref{FNCSignalLaplaceTransTk} is provided in Appendix \ref{Appendix:LaplaceTransInFNC}.
\end{IEEEproof}
\begin{theorem}
\label{Lemma:LaplaceTransInFNCStep2}
$\mathcal{L}_{I^{M,\mathrm{FN}}|r}\left(z\right)$ is calculated by
\begin{subequations}
\begin{align}
\label{FNCInterferenceLaplaceTrans}
\mathcal{L}_{I^{M,\mathrm{FN}}|r}\left(z\right)& = \mathcal{L}_{I_1^{M,\mathrm{FN}}|r}\left(z\right)\mathcal{L}_{I_2^{M,\mathrm{FN}}|r}
\left(z\right), \\
\label{At_rInterferenceLaplaceTransFNC}
\mathcal{L}_{I_1^{M,\mathrm{FN}}|r}\left(z\right) &=
\sum_{\mu\in \{M,S\}}\sum_{\nu\in \{L,N\}}
\frac{p_{\mu} p_{\nu}\mathbb{I}\left(r\leq D\right)}{\left(1+\frac{1}{N_\nu}za_{\mu,\nu}r^{-\alpha_{\nu}}\right)^{N_\nu}}+\sum_{\mu\in \{M,S\}}
\frac{p_{\mu}\mathbb{I}\left(r>D\right)}{\left(1+\frac{1}{N_N}za_{\mu,N}r^{-\alpha_{N}}\right)^{N_N}},\\
\label{Outof_rInterferenceLaplaceTransFNC}
\mathcal{L}_{I_2^{M,\mathrm{FN}}|r}\left(z\right)
&= \exp\left\{-\pi\lambda_B \sum_{\mu \in \{M,S\}} p_{\mu}\left(\Xi_{\mu}\left(z,\mathrm{max}\left(D,r\right)\right)
+ \mathbb{I}\left(r\leq D\right)\sum_{\nu \in \{L,N\}}p_{\nu}\Theta_{\mu,\nu}\left(z,r\right)\right) \right\}
\end{align}
\end{subequations}
where $I_1^{M,\mathrm{FN}}$ and $I_2^{M,\mathrm{FN}}$ are the interfering power from the $(M+1)^{\mathrm{th}}$ nearest BS and the BSs outside the region $\mathcal{B}\left(o,r\right)$, respectively, and
\begin{subequations}
\begin{align}
\Xi_{\mu}\left(z,x\right)&\triangleq \int_x^{+\infty}2t\left( 1 - \mathcal{L}_{h^{(N)}}\left(za_{\mu,N}t^{-\alpha_N}\right) \right)\mathrm{d}t
\nonumber\\
&=\beta_\mu\left(x,z\right)+
\frac{a_{\mu,N}\alpha_N  z x^{2-\alpha_N}}{\alpha_N-2}{}_2F_{1}\left(1 - \frac{2}{\alpha_N} , 1 + N_N ;
  2 - \frac{2}{\alpha_N}; -\frac{a_{\mu,N}z}{N_N x^{\alpha_N}}\right), \\
\Theta_{\mu,\nu}\left(z,x\right) &\triangleq
D^2 - x^2 -\Lambda_{\mu,\nu}\left(z,D,x\right), \label{NakTheataFunction}\\
\beta_\mu\left(x,z\right)&\triangleq \left.\left(x^2 - \left(1+\frac{za_{\mu,N}}{N_N}x^{-\alpha_N}\right)^{N_N}\right)\middle/\left(1+\frac{za_{\mu,N}}{N_N}x^{-\alpha_N}\right)^{N_N}\right.,
\end{align}
\end{subequations}
\end{theorem}
\begin{IEEEproof}
When $D_{M+1}$ is fixed as $r$, $I_1^{M,\mathrm{FN}}$ and $I_2^{M,\mathrm{FN}}$ are mutually independent, therefore we obtain \eqref{FNCInterferenceLaplaceTrans}.
The derivations of \eqref{At_rInterferenceLaplaceTransFNC} and \eqref{Outof_rInterferenceLaplaceTransFNC} are provided in Appendix \ref{Appendix:LaplaceTransInFNC}.
\end{IEEEproof}

Inserting \eqref{FNCSignalLaplaceTrans} and \eqref{FNCInterferenceLaplaceTrans} into \eqref{AverageRateFNC}, we can calculate the average rate of the typical cell-edge user in FNC systems. Note that in \eqref{AverageRateFNC}, by making a change of variable, i.e., $\lambda_B\pi \left(r^2-D_E^2\right)\rightarrow x$, \eqref{AverageRateFNC} can be reformulated into the form of $\int_0^{+\infty} g(x)\exp(-x)\mathrm{d}x$, which can be effective calculated by using the Gauss-Laguerre quadrature \cite{P.J.Davis}.

\subsection{Average rate of the typical cell-edge user in FRC scheme}
The following theorem presents the average rate of the typical cell-edge user in FRC scheme.
\begin{theorem}
$\bar{R}^{D_{CO},\mathrm{FR}}$ is given by
\begin{align}
\bar{R}^{D_{CO},\mathrm{FR}}& =
\mathbb{E}_{T^{D_{CO},\mathrm{FR}},I^{D_{CO},\mathrm{FR}}}\left(\ln\left(1 + \frac{T^{D_{CO},\mathrm{FR}}}{I^{D_{CO},\mathrm{FR}} + N_0}\right)\right)\nonumber\\
& =\int_{0}^{+\infty}\left( 1- \mathcal{L}_{T^{D_{CO},\mathrm{FR}}}\left(z\right)\right) \mathcal{L}_{I^{D_{CO},\mathrm{FR}}}\left(z\right) \exp \left( -N_0z \right)\frac{\mathrm{d}z}{z}.\label{AverageRateFRC}
\end{align}
\end{theorem}
\begin{IEEEproof}
Due to the fact that $T^{D_{CO},\mathrm{FR}}$ and $I^{D_{CO},\mathrm{FR}}$ are  independent RVs. By using Lemma \ref{LemmaRate}, we obtain \eqref{AverageRateFRC}.
\end{IEEEproof}

To calculate \eqref{AverageRateFRC}, we need the LT of the signal power $T^{D_{CO},\mathrm{FR}}$ and the interfering power $I^{D_{CO},\mathrm{FR}}$, which are provided in the following theorem.
\begin{theorem} \label{Theorem:LaplaceTransInFRC}
$\mathcal{L}_{T^{D_{CO},\mathrm{FR}}}\left(z\right)$ and $\mathcal{L}_{I^{D_{CO},\mathrm{FR}}}\left(z\right)$ are given by
\begin{subequations}
\label{Eqn:LaplaceTransInFRC}
\begin{align}
\mathcal{L}_{T^{D_{CO},\mathrm{FR}}}\left(z\right)&=\exp\Bigg\{-\lambda_B\pi \left(D_{CO}^2-D_E^2\right)+\lambda_B\pi
\Bigg\{\sum_{\nu\in \{L,N\}} p_\nu \Lambda_{M,\nu} \left( z,\mathrm{min}\left(D_{CO},D\right),D_E\right)\nonumber \\
&\quad\quad\quad\quad\quad\quad\quad\quad\quad\quad\quad\ \ \
+\mathbb{I}\left(D_{CO}> D\right)\Lambda_{M,N} \left( z,D_{CO},D\right)\Bigg\}\Bigg\} ,
\label{FRCSignalLaplaceTrans} \\
\mathcal{L}_{I^{D_{CO},\mathrm{FR}}}\left(z\right)&=\mathcal{L}_{I_2^{M,\mathrm{FN}}|r}
\left(z\right)\big |_{r=D_{CO}},\label{FRCInterferenceLaplaceTrans}
\end{align}
\end{subequations}
where $\Lambda_{\mu,\nu}\left(z,x,y\right)$ and $\mathcal{L}_{I_2^{M,\mathrm{FN}}|r}
\left(z\right)$ are defined in \eqref{LambdaFunc} and \eqref{Outof_rInterferenceLaplaceTransFNC}, respectively.
\end{theorem}
\begin{IEEEproof}
The proof is given in Appendix \ref{Appendix:LaplaceTransInFRC}.
\end{IEEEproof}

Substituting \eqref{Eqn:LaplaceTransInFRC} into \eqref{AverageRateFRC}, we obtain the average rates of the typical cell-edge user in FRC systems.

\subsection{Numerical Example}

%\begin{figure}
%    \centering
%  \subfigure[Average rates comparisons of the analytical results and the simulation results.]{
%    \label{RatevsM:sub1} %% label for first subfigure
%    \includegraphics[width=3 in]{RvsM.eps}}
%    \hspace{.1in}
%  \subfigure[The time consumptions to obtain the analytical and simulation results in (a).]{
%    \label{RatevsM:sub2} %% label for second subfigure
%    \includegraphics[width=3 in]{RvsMTime.eps}}
%  \caption{Average rate of the typical cell-edge user. System parameters are given by $\alpha_L\left(\alpha_N\right)=2\left(2.92\right)$, $C_L\left(C_N\right)=-61.4\left(-72\right)~\mathrm{dB}$ \cite{T.Pappaport2013},
%$G_M\left(G_S\right)=15(-3)~\mathrm{dB}$, $N_L\left(N_N\right)=3\left(1\right)$, $\mathrm{BW}=1~\mathrm{GHz}$, $F=5~\mathrm{dBm}$, $\theta_T=15^{\circ}$,
%$p_L=0.11$, $P_T=20~\mathrm{dBm}$, $\rho=90~(m)$, and $\chi =  1$.}
%  \label{RatevsMTot} %% label for entire figure
%
%\end{figure}

\begin{figure}
    \centering
  \subfigure[Average rate in FNC scheme.]{
    \label{RATE1:FNC} %% label for first subfigure
    \includegraphics[width=2.7 in]{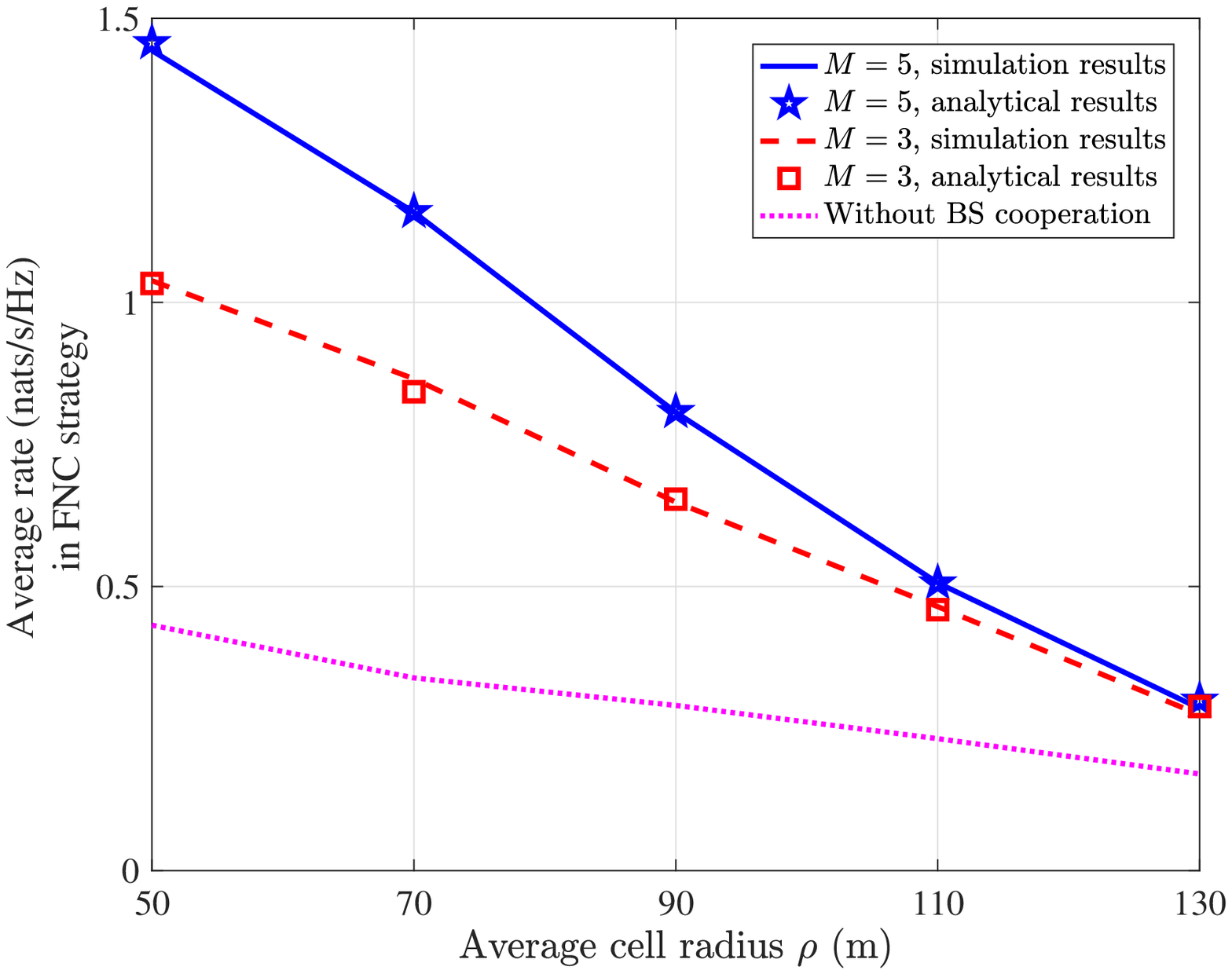}}
    \hspace{.1in}
  \subfigure[Average rate in FRC scheme.]{
    \label{RATE1:FRC} %% label for second subfigure
    \includegraphics[width=2.7 in]{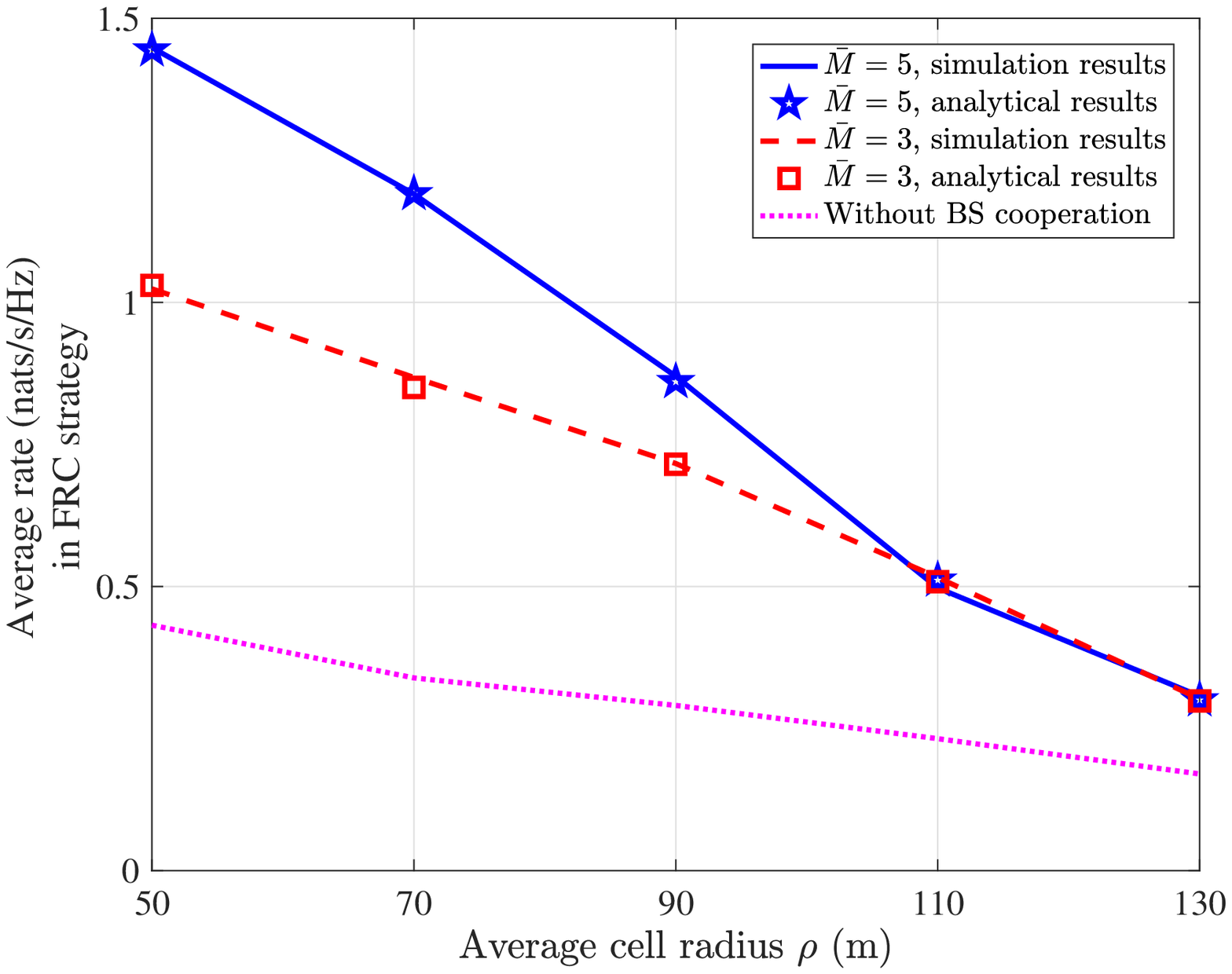}}
  \caption{Average rate of the typical cell-edge user versus the average cell radius.}
  \label{RATE1} %% label for entire figure
\end{figure}
In this subsection, we provide some numerical examples to verify the analytical results of the average rate given in Theorem 1 and Theorem 4.
In Fig. \ref{RATE1}, we plot the average rates against the average cell radiuses, i.e., $\rho$.
The system settings are given by $\alpha_L\left(\alpha_N\right)=2\left(2.92\right)$, $C_L\left(C_N\right)=-61.4\left(-72\right)~\mathrm{dB}$ \cite{T.Pappaport2013},
$G_M\left(G_S\right)=15(-3)~\mathrm{dB}$, $N_L\left(N_N\right)=3\left(1\right)$, $\mathrm{BW}=1~\mathrm{GHz}$, $F=5~\mathrm{dBm}$, $\theta_T=15^{\circ}$,
$p_L=0.11$, $P_T=20~\mathrm{dBm}$, and $\chi =  1$.
Note that we use $M$ to denote the number of the cooperative BSs in FNC scheme, and for notational convenience, we use $\bar{M}$ to equivalently denote the radius of the cooperative region, i.e., $D_{CO}$, in FRC scheme by setting $\bar{M} = \pi\lambda_B\left(D_{CO}^2-D_E^2\right)$.
%\footnote{
%It is easy to prove that in this case, the average number of cooperative BSs in FRC scheme is $\bar{M}$.
%Using $\bar{M}$ to equivalently represent $D_{CO}$ provides us a simple yet fair way to compare the performance between the FNC and FRC strategies.
%}
The simulation results here are obtained by averaging across $10^4$ random BS deployments
following the detailed steps given in Section VI.
In Fig. \ref{RATE1}, the performance for the non-cooperative cases where the typical edge user only connects to the nearest BS are also provided for comparison.
From Fig. \ref{RATE1}, we can find that the analytical results are very close
to the simulation results, which validates our analytical expressions for the average rate.
Besides, we observe that compared with the non-cooperative cases, BS cooperation brings an enormous improvement in term of the average rate.
Under the dense BS deployment, for example, $\rho = 50$ (m), both FNC and FRC strategies achieve significant increases in the average rate, i.e., from $0.43$ to $1.46$ and $1.44$ nats/s/Hz, respectively, when $M(\bar{M})=5$.
Under the less dense BS deployments, the growth of the average rates can still be noticed.
For example, when $\rho=130$ (m), the average rates are increased by around $60\%$  under both the FNC and FRC strategies, i.e., from about $0.18$
to $0.29$ nats/s/Hz.
The basic principle behind this phenomenon is that the BS cooperation transforms the potential strong interferers into the cooperative BSs and therefore not only
strengthens the signal power but also reduces the interfering power.

\section{Outage Probability of the Typical Cell Edge User}
In this section, the outage probabilities of the typical cell-edge user are investigated.
We first present the exact expressions for the outage probabilities for both FNC and FRC schemes in the following two theorems.

\subsection{Exact analytical results}
\begin{theorem}
The outage probabilities of the typical cell-edge user in FNC mmWave cellular networks are given by
\begin{align}
\label{OutageFNC}
    \mathcal{P}_{O}^{M,\mathrm{FN}}\left(\tau\right) &=\frac{1}{2}+ \int_{0}^{+\infty}\int_{-\infty}^{+\infty} f_{M+1}\left(r\right) \mathrm{Im}\left\{\mathcal{L}_{T^{M,\mathrm{FN}}|r}\left(\mathrm{j}\omega\right) \mathcal{L}_{I^{M,\mathrm{FN}}|r}\left(-\mathrm{j}\omega \tau\right)\exp \left(\mathrm{j}\omega N_0 \tau \right)\right\}
    \frac{\mathrm{d}\omega}{\pi\omega}\mathrm{d}r,
\end{align}
where $\mathcal{L}_{T^{M,\mathrm{FN}}|r}\left(-\mathrm{j}\omega\right)$ and $\mathcal{L}_{I^{M,\mathrm{FN}}|r}\left(\mathrm{j}\omega\tau\right)$ are provided in Theorem  \ref{Lemma:LaplaceTransInFNCStep1} and \ref{Lemma:LaplaceTransInFNCStep2}, respectively.
\end{theorem}
\begin{IEEEproof}
We have
\begin{align}
\mathcal{P}_{O}^{M,\mathrm{FN}}\left(\tau\right)= \mathbb{P}\left\{T^{M,\mathrm{FN}} -\tau I^{M,\mathrm{FN}} \leq \tau N_0\right\} = \mathbb{E}_{D_{M+1}}\left[
\mathbb{P}\left\{\varpi^{M,\mathrm{FN}} \leq \tau N_0 | D_{M+1}=r \right\}
\right],
\label{OutageFNCproof}
\end{align}
where $\varpi^{M,\mathrm{FN}}\triangleq T^{M,\mathrm{FN}} -\tau I^{M,\mathrm{FN}}$. Using the Inverse Theorem \cite{J.Gil-Pelaez1951}, we have
\begin{align}
\mathbb{P}\left\{\varpi^{M,\mathrm{FN}} \leq \tau N_0 | D_{M+1}=r \right\}=\frac{1}{2}+\int_{0}^{+\infty}
\mathrm{Im}\left\{\exp\left(\mathrm{j}\omega \tau N_0\right)
\mathcal{L}_{\varpi^{M,\mathrm{FN}}|r}\left(\mathrm{j}\omega\right)\right\}
\frac{\mathrm{d}\omega}{\pi\omega},
\label{InverseTheorem}
\end{align}
where $\mathcal{L}_{\varpi^{M,\mathrm{FN}}|r}\left(\mathrm{j}\omega\right) = \mathcal{L}_{T^{M,\mathrm{FN}}|r}\left(\mathrm{j}\omega\right)
\mathcal{L}_{I^{M,\mathrm{FN}}|r}\left(-\mathrm{j}\tau \omega\right)$. Inserting \eqref{InverseTheorem} into \eqref{OutageFNCproof}, we obtain \eqref{OutageFNC}.
\end{IEEEproof}

\begin{theorem}
The outage probabilities of the typical cell-edge user in FRC mmWave cellular networks are given by
\begin{align}
\label{OutageFRC}
    \mathcal{P}_{O}^{D_{CO},\mathrm{FR}}\left(\tau\right) &= \frac{1}{2}+\int_{0}^{+\infty}  \mathrm{Im}\{\mathcal{L}_{T^{D_{CO},\mathrm{FR}}}\left(\mathrm{j}\omega\right)
    \mathcal{L}_{I^{D_{CO},\mathrm{FR}}}\left(-\mathrm{j}\omega \tau\right)\exp \left(\mathrm{j}\omega N_0 \tau \right)\}\frac{\mathrm{d}\omega}{ \pi\omega },
\end{align}
where $\mathcal{L}_{T^{D_{CO},\mathrm{FR}}}\left(\mathrm{j}\omega\right)$ and $\mathcal{L}_{I^{D_{CO},\mathrm{FR}}}\left(-\mathrm{j}\omega\tau\right)$ are provided in Theorem \ref{Theorem:LaplaceTransInFRC}.
\end{theorem}
\begin{IEEEproof}
We first have$\mathcal{P}_{O}^{D_{CO},\mathrm{FR}}\left(\tau\right)=\mathbb{P}\left\{\varpi^{D_{CO},\mathrm{FR}} \leq \tau N_0  \right\}$
with $\varpi^{D_{CO},\mathrm{FR}} \triangleq T^{D_{CO},\mathrm{FR}} - \tau I^{D_{CO},\mathrm{FR}}$. Then, using the Inverse Theorem \cite{J.Gil-Pelaez1951}, we directly obtain \eqref{OutageFRC}.
\end{IEEEproof}

\begin{figure}
    \centering
  \subfigure[Outage probabilities comparison between the analytical results and the simulation results.]{
    \label{OutvsThre:sub1} %% label for first subfigure
    \includegraphics[width=2.7 in]{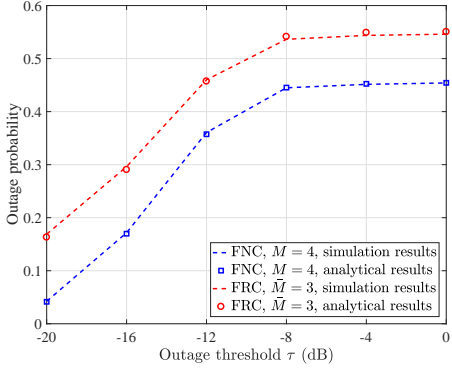}}
    \hspace{.1in}
  \subfigure[The time consumption to obtain the analytical and simulation results in (a).]{
    \label{OutvsThre:sub2} %% label for second subfigure
    \includegraphics[width=2.7 in]{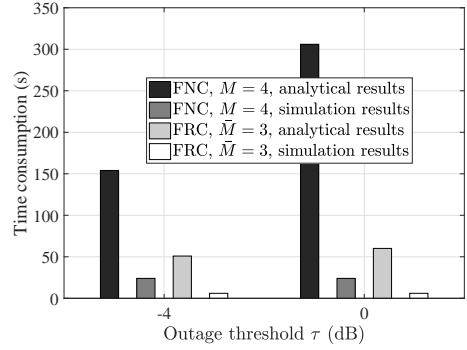}}
  \caption{Outage probability of the typical cell-edge user.}
  \label{OutvsThreTot} %% label for entire figure
\end{figure}

With \eqref{OutageFNC} and \eqref{OutageFRC}, we can compute the exact outage probabilities of the typical cell-edge user in FNC and FRC mmWave cellular networks, respectively. We validate the analytical results in \eqref{OutageFNC} and \eqref{OutageFRC} in Fig. \ref{OutvsThreTot}, where we set $p_L=0.2$ and other the system parameters are set the same as that in Fig. \ref{RATE1}.
As Fig. \ref{OutvsThre:sub1} shows, the analytical outage probabilities are very close to the simulation results which means that the analytical results are valid. However,
in Fig. \ref{OutvsThre:sub2}, we plot the time consumptions that are required to calculate the analytical results and the simulation results, respectively.
As we can see, to obtain the analytical results, we need to spend more computing time than to directly carry out the simulation, which limits the application of the analytical results in \eqref{OutageFNC} and \eqref{OutageFRC}. We also have to note that the calculation of the outage probabilities in FNC system are more time-consuming than those in FRC system, because \eqref{OutageFNC} have a twofold integral while there is only a single integral in \eqref{OutageFRC}.

In the following,
we attempt to provide computationally efficient approximations to these two outage probabilities.
The basic principles of our approximation method are summarized as follows.
\begin{enumerate}
\item Field measurement data reveal that compared with the thermal noise, the interference in mmWave band is not dominant, especially for downlink transmissions \cite{S.Rangan2014,M.R.Akdeniz2014}.
    In fact, mmWave signal is greatly suffered by the blockage effect, which causes that the strong interference sources usually only exist when they are close to the receiver.
    However, the BS cooperation transforms the potential strong interfering BSs into the cooperative BSs, and the interfering BSs are now kept far away from the receiver.
    Besides, the directional transmissions provides high array gain at the intended user and greatly suppresses the side lobe interfering power.
    Based on these observations, we assume that the interfering power can be neglected;
\item By neglecting the interference, the outage probability depends on the desired signal power $T$ and the noise power $N_0$. We observe that for fixed BS deployment, $T$ is the summation of several Gamma RVs, and though its CDF can not be obtained in a simple form, we can accurately approximate it by using the technique of Gamma approximation \cite{R.W.Heath2013}.
\end{enumerate}

\subsection{An approximation of the outage probability in FNC scheme}
First, we divide the desire signal power $T^{M,FN}$ into two RVs as follow
\begin{align}
\label{InitailAppFNC}
T^{M,FN}=\begin{cases}
T_L^{M,\mathrm{FN}}, & \mathbb{P}\left\{T^{M,\mathrm{FN}}=T_L^{M,\mathrm{FN}}\right\} = 1 - p_{NL},\\
T_N^{M,\mathrm{FN}}, & \mathbb{P}\left\{T^{M,\mathrm{FN}}=T_N^{M,\mathrm{FN}}\right\} = p_{NL},\\
\end{cases}
\end{align}
where $p_{NL}$ denotes the probability that all the cooperative BSs are in NLOS state, which is calculated as
$p_{NL} = \int_{D_E}^{D}f_{M+1}\left(r\right)\left(1-p_L\right)^M\mathrm{d}r + \int_{D}^{+\infty}f_{M+1}\left(r\right)\left(1-p_L\frac{D^2 - D_E^2}{r^2-D_E^2}\right)^M\mathrm{d}r$,
$T_L^{M,\mathrm{FN}}$ and $T_N^{M,\mathrm{FN}}$ denote the signal powers when there is at least one cooperative BS is in LOS state  and
when all the cooperative BSs are in NLOS state, respectively.
Both $T_L^{M,\mathrm{FN}}$ and $T_N^{M,\mathrm{FN}}$ are RVs, and obviously, they are mutually exclusive and exhaustive.
Now, we use two Gamma RVs to approximate $T_L^{M,\mathrm{FN}}$ and $T_N^{M,\mathrm{FN}}$, respectively.
\subsubsection{Approximation of $T_L^{M,\mathrm{FN}}$}
Due to the fact that the signal power from NLOS links suffer much more severe attenuation than that from LOS links.
Therefore, when LOS links exist, we neglect the signal power from the cooperative BSs in NLOS state. Then, we can approximate $T_L^{M,\mathrm{FN}}$ as
\begin{align}
T_L^{M,\mathrm{FN}}\approx \breve{T}_L^{M,\mathrm{FN}} = \sum_{i\in \Omega_{L}^{M,\mathrm{FN}}} a_{M,L}h_{i}^{(L)}r_{i}^{-\alpha_L},
\label{APPTFNC1}
\end{align}
where $\Omega_{L}^{M,\mathrm{FN}}$ is the set of the cooperative BSs that are in LOS state.
Denote $S_L^{M,\mathrm{FN}}$ the cardinality of $\Omega_{L}^{M,\mathrm{FN}}$, i.e., $S_L^{M,\mathrm{FN}}\triangleq |\Omega_{L}^{M,\mathrm{FN}}|$, then
$S_L^{M,\mathrm{FN}}$ is a RV. Conditioning on $D_{M+1}=r$, the conditional PMF of $S_L^{M,\mathrm{FN}}$ is given by
\begin{align}
\mathcal{P}_{S_L^{M,\mathrm{FN}}}\left(m,r\right)
&\triangleq\mathbb{P}\left\{S_L^{M,\mathrm{FN}}=m\Big|S_L^{M,\mathrm{FN}}\geq1,D_{M+1}=r\right\} \nonumber\\
&= \mathbb{I}\left\{r\leq D\right\}
\begin{pmatrix}
M\\
m
\end{pmatrix}
\frac{1}{p_{NL}}
p_L^m\left(1-p_L\right)^{M-m}\nonumber\\
&\quad + \mathbb{I}\left\{r > D\right\}
\begin{pmatrix}
M\\
m
\end{pmatrix}
\frac{1}{p_{NL}}
\left(\frac{D^2 -D_E^2 }{r^2 - D_E^2}p_L\right)^m\left(1-\frac{D^2 -D_E^2 }{r^2 - D_E^2}p_L\right)^{M-m},
\label{PMFNumLOS}
\end{align}
where $m = 1,2,\cdot\cdot\cdot,M$.
Based on \eqref{APPTFNC1} and \eqref{PMFNumLOS}, we now use a Gamma RV to approximate $\tilde{T}_L^{M,FN}$ in the following Proposition.
\begin{proposition}
$\breve{T}_L^{M,FN}$ can be approximated by $
\breve{T}_L^{M,FN} \approx \tilde{T}_L^{M,FN}\sim \mathrm{Gamma}\left(\kappa_L^{M,FN}, \rho_L^{M,FN}\right)$
where $\kappa_L^{M,\mathrm{FN}}$ and $\rho_L^{M,\mathrm{FN}}$ are given by
\begin{align}
\kappa_L^{M,FN}= \frac{\left(\mathbb{E}\left(\breve{T}_L^{M,FN}\big|S_L^{M,FN}\geq1\right)\right)^2}
{\mathbb{D}\left(\breve{T}_L^{M,FN}\big|S_L^{M,FN}\geq1\right)}, \quad
\rho_L^{M,FN} = \frac{\mathbb{E}\left(\breve{T}_L^{M,FN}\big|S_L^{M,FN}\right)}{\kappa_L^{M,FN}}.
\label{GammaApproximate1}
\end{align}
\end{proposition}
\begin{IEEEproof}
$\kappa_L^{M,\mathrm{FN}}$ and $\rho_L^{M,\mathrm{FN}}$ are obtained by matching the first and second order moments of $\tilde{T}_L^{M,\mathrm{FN}}$ and $\breve{T}_L^{M,FN}$, and the expressions of $\kappa_L^{M,\mathrm{FN}}$ and $\rho_L^{M,\mathrm{FN}}$ are provided in Appendix \ref{APP:3}.
\end{IEEEproof}

\subsubsection{Approximation of $T_N^{M,\mathrm{FN}}$}
In this case, all the cooperative BSs are in NLOS state, and
we have $T_N^{M,\mathrm{FN}}=\sum_{i\in \Omega^{M,\mathrm{FN}}} a_{M,N}h_i^{(N)}r_i^{-\alpha_N}$.
Note that the cardinality of $\Omega^{M,\mathrm{FN}}$ is fixed as $M$. We then have the following Proposition.
\begin{proposition}
$T_N^{M,\mathrm{FN}}$ can be approximated by $
T_N^{M,\mathrm{FN}} \approx \tilde{T}_N^{M,\mathrm{FN}}\sim \mathrm{Gamma}\left(\kappa_N^{M,\mathrm{FN}}, \rho_N^{M,\mathrm{FN}}\right)$
where $\kappa_L^{M,\mathrm{FN}}$ and $\rho_L^{M,\mathrm{FN}}$ are obtained by matching the first and second order moments, which are given by
\begin{align}
\kappa_N^{M,\mathrm{FN}}= \frac{\left(\mathbb{E}\left(T_N^{M,\mathrm{FN}}\right)\right)^2}{\mathbb{D}\left(T_N^{M,\mathrm{FN}}\right)}, \quad
\rho_N^{M,\mathrm{FN}} = \frac{\mathbb{E}\left(T_N^{M,\mathrm{FN}}\right)}{\kappa_N^{M,\mathrm{FN}}}.
\label{GammaApproximate2}
\end{align}
\end{proposition}
\begin{IEEEproof}
$\kappa_N^{M,\mathrm{FN}}$ and $\rho_N^{M,\mathrm{FN}}$ are obtained by matching the first and second order moments of $T_N^{M,\mathrm{FN}}$ and $\tilde{T}_N^{M,\mathrm{FN}}$, and the expressions of $\kappa_N^{M,\mathrm{FN}}$ and $\rho_N^{M,\mathrm{FN}}$ are provided in Appendix \ref{APP:3}.
\end{IEEEproof}

Based on \eqref{InitailAppFNC} and Proposition 1 and 2, we can approximate the outage probability of FNC systems as
\begin{align}
\mathcal{P}_{O}^{M,\mathrm{FN}}\left(\tau\right) \approx \left(1 - p_{NL}\right)
F_{\tilde{T}_L^{M,FN}}\left(\tau N_0\right) + p_{NL}F_{\tilde{T}_N^{M,FN}}\left(\tau N_0\right),
\label{FinalAppFNC}
\end{align}
where $F_{\tilde{T}_\nu^{M,\mathrm{FN}}}\left( t \right)$ is the CDF of $\tilde{T}_\nu^{M,\mathrm{FN}}$ for $\nu \in \{L,N\}$, i.e.,
\begin{align}
F_{\tilde{T}_\nu^{M,\mathrm{FN}}}\left( t \right) = \gamma\left(\kappa_\nu^{M,\mathrm{FN}},t\big/\rho_\nu^{M,\mathrm{FN}}\right)\bigg/\Gamma\left(\kappa_\nu^{M,\mathrm{FN}}\right).
\end{align}

Note that \eqref{FinalAppFNC} is computationally much more efficient than \eqref{OutageFNC} because \eqref{FinalAppFNC} only involves several Gamma functions while \eqref{OutageFNC} requires to numerically calculate a two-fold integral.
We will show the accuracy of the approximation in \eqref{FinalAppFNC} in the numeric section in Section VI.

\subsection{An approximation of the outage probability in FRC scheme}
We approximate the outage probabilities of the typical cell-edge user in FRC systems in this subsection.
Without loss of generality, we assume $D_{CO}\leq D$ in this subsection. The corresponding results can be easily extended to the opposite case.
Under the FRC scheme, the signal power $T^{D_{CO},FR}$ can be written as the summation of two independent RVs as follows
\begin{align}
T^{D_{CO},\mathrm{FR}}  =  T_{L}^{D_{CO},FR} + T_{N}^{D_{CO},\mathrm{FR}},
\label{InitailAPPFRC}
\end{align}
where $T_{L}^{D_{CO},\mathrm{FR}}\triangleq\sum_{i\in \Omega_{L}^{D_{CO},\mathrm{FR}}} a_{M,L}h_i^{(L)}r_{i}^{-\alpha_L} $ and
$T_{N}^{D_{CO},\mathrm{FR}}\triangleq\sum_{i\in \Omega_{N}^{D_{CO},\mathrm{FR}}} a_{M,N}h_i^{(N)}r_{i}^{-\alpha_N}$ are the sum signal power from the LOS BSs and the NLOS BSs within the cooperative region, respectively,  and $\Omega_{L}^{D_{CO},\mathrm{FR}}$ and $\Omega_{N}^{D_{CO},\mathrm{FR}}$ are the sets of the cooperative BSs that are in LOS and NLOS state, respectively.
Defining $S_\nu^{D_{CO},\mathrm{\mathrm{FR}}}=\left|\Omega_{\nu}^{D_{CO},\mathrm{FR}}\right|$ for $\nu\in\{L,N\}$, we have
\begin{align}
T_{\nu}^{D_{CO},\mathrm{FR}}=
\begin{cases}
0,&\mathbb{P}\left\{T_{\nu}^{D_{CO},\mathrm{FR}}=0\right\}=p_{0,\nu},\\
\breve{T}_{\nu}^{D_{CO},FR},& \mathbb{P}\left\{T_{\nu}^{D_{CO},FR}=\breve{T}_{\nu}^{D_{CO},FR}\right\}=1-p_{0,\nu},
\end{cases}
\end{align}
where for $\nu\in\{L,N\}$, $\breve{T}_{\nu}^{D_{CO},\mathrm{FR}}$ is defined as the signal power from the cooperative BSs in $\Omega_{\nu}^{D_{CO},\mathrm{FR}}$ conditioning on $S_\nu^{D_{CO},\mathrm{FR}} \geq 1$, and $p_{0,\nu}^{D_{CO},\mathrm{FR}}$ is the probability that there is no BS in $\Omega_{\nu}^{D_{CO},\mathrm{FR}}$,
which is given by
\begin{align}
p_{0,\nu}\triangleq\mathbb{P}\left\{S_\nu^{D_{CO},\mathrm{FR}} = 0\right\}=\exp\left(-p_\nu\lambda_B\left(D_{CO}^2-D_E^2\right)\right).
\end{align}
Now, we use two RVs to approximate $\breve{T}_{\nu}^{D_{CO},\mathrm{FR}}$ for $\nu\in\{L,N\}$.
\begin{proposition}
$\breve{T}_{\nu}^{D_{CO},\mathrm{FR}}$ for $\nu\in\{L,N\}$ can be approximated by $
\breve{T}_{\nu}^{D_{CO},\mathrm{FR}}\approx
\tilde{T}_{\nu}^{D_{CO},\mathrm{FR}} \sim \mathrm{Gamma}\left( \kappa_{\nu}^{D_{CO},\mathrm{FR}}, \rho_{\nu}^{D_{CO},\mathrm{FR}} \right)$
where
\begin{align}
\kappa_{\nu}^{D_{CO},\mathrm{FR}}=\frac{\left(\mathbb{E}\left(\breve{T}_{\nu}^{D_{CO},FR}\big||\Omega_{\nu}^{D_{CO},\mathrm{FR}}|>0\right)\right)^2}
{\mathbb{D}\left(\breve{T}_{\nu}^{D_{CO},FR}\big| |\Omega_{\nu}^{D_{CO},\mathrm{FR}} |>0\right)}, \quad
\rho_{\nu}^{D_{CO},\mathrm{FR}}=
\frac{\mathbb{E}\left(\breve{T}_{\nu}^{D_{CO},\mathrm{FR}}\big||\Omega_{\nu}^{D_{CO},\mathrm{FR}}|>0\right)}{\kappa_{\nu}^{D_{CO},\mathrm{FR}}}.
\label{GammaApproximate3}
\end{align}
\end{proposition}
\begin{IEEEproof}
The detailed expressions of $\kappa_{\nu}^{D_{CO},\mathrm{FR}}$ and $\rho_{\nu}^{D_{CO},\mathrm{FR}}$ for $\nu\in\{L,N\}$ are provided in the Appendix \ref{APP:3}.
\end{IEEEproof}

Based on \eqref{InitailAPPFRC} and Proposition 3, we can approximate the outage probability of the typical cell-edge user in FRC systems as
\begin{align}
\mathcal{P}_{O}^{D_{CO},\mathrm{FR}}\left(\tau\right)& \approx
\mathbb{P}\left\{T_{L}^{D_{CO},\mathrm{FR}} + T_{N}^{D_{CO},\mathrm{FR}} \leq \tau N_0\right\}
\nonumber \\
& \approx p_{0,L}p_{0,N}+
\left(1-p_{0,L}\right)p_{0,N}F_{\tilde{T}_{L}^{D_{CO},\mathrm{FR}}}\left(\tau N_0\right)+
p_{0,L}\left(1-p_{0,N}\right)F_{\tilde{T}_{N}^{D_{CO},\mathrm{FR}}}\left(\tau N_0\right)\nonumber\\
&\quad +
\left(1-p_{0,L}\right)\left(1-p_{0,N}\right)
\int_0^{\tau N_0}f_{\tilde{T}_{L}^{D_{CO},\mathrm{FR}}}\left(t\right)F_{\tilde{T}_{N}^{D_{CO},\mathrm{FR}}}\left(\tau N_0 - t \right)\mathrm{d}t,
\label{FinalAppFRC}
\end{align}
where $f_{\tilde{T}_{L}^{D_{CO},\mathrm{FR}}}\left(x\right)$ is the PDF of $\tilde{T}_{L}^{D_{CO},\mathrm{FR}}$ and $F_{\tilde{T}_{N}^{D_{CO},\mathrm{FR}}}\left(x\right)$ are the CDF of $\tilde{T}_{N}^{D_{CO},\mathrm{FR}}$.
Note that both of \eqref{OutageFRC} and \eqref{FinalAppFRC} involve integral operations.
However, the integrand in \eqref{OutageFRC} is oscillating and involves hypergeometric function in the complex plane, the calculation of
which is much more complicated than that in \eqref{FinalAppFRC}.
The accuracy and the efficiency of the approximations in \eqref{FinalAppFNC} and \eqref{FinalAppFRC} will be show in Fig. \ref{Out1} in next section.

\section{Simulation Results \& Discussion}
In this section, simulation results are provided to evaluate the performance of the typical edge
user in the BS cooperation-aided mmWave networks.
Unless specified, the simulation parameters are the same as that given in Fig. \ref{RATE1}.
Note that to illustrate the performance of the cell-edge users, we set $\chi=1$, i.e., $D_E=\rho$, which means that there is no BS located within the distance of average cell radius around the typical cell-edge user.

All the simulation results are obtained by carrying out the following steps: (a) we set an edge user at the origin; (b) randomly generate the locations of the BSs in the annular region $\mathcal{R}\left(D_E,D_\infty\right)$, where $D_\infty$ is large enough to eliminate the impact of the BSs outside $\mathcal{B}\left(o,D_\infty\right)$; (c) determine the cooperative BSs according to the cooperative strategies in Section \ref{Section:CooperativeStrategy}; (d) the states (LOS or NLOS) and the small scale power fading factors from each BS to the typical edge user are randomly generated according to Section \ref{Section2}, and for each interfering BS, the angle of departure is randomly generated to determine the array gain; (e) record the signal power and the interfering power according to \eqref{SignalInterference}, and return to step (a) until $10^4$ times of trails are finished. In our simulation, we set $D_\infty=2000$ (m).
The simulation results are presented in the following subsections.

\subsection{Average Rate of the Typical Cell-Edge User}
\begin{figure}[tp]
\begin{center}
\includegraphics[width=2.7 in]{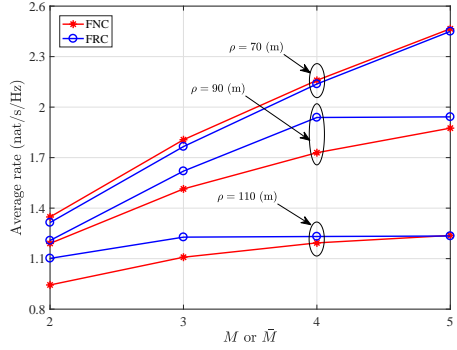}
\caption{Average rates versus the number of cooperative BSs.}\label{ARM}
\end{center}
\vspace{-5mm}
\end{figure}

In Fig. \ref{ARM}, we plot the average rates in FNC and FRC schemes versus $M$ and $\bar{M}$, respectively, where we set $N_L = 4$, $N_N = 1$, and $p_L = 0.2$.
Note that the results for the FNC and FRC strategies in Fig. \ref{ARM} are obtained by calculating \eqref{AverageRateFNC} and \eqref{AverageRateFRC}, respectively, without neglecting the interference.
From Fig. \ref{ARM}, we observe that under the condition that $\bar{M}=M $, the FRC scheme achieves higher average rates than the FNC scheme when the BSs get sparse (e.g., when $\rho = 90,100$ (m)), but when the BSs are dense (e.g., when $\rho = 70$ (m)), the FNC scheme outperforms the FRC scheme. This can be interpreted as follows. In FRC scheme, the number of cooperative BSs is a Poisson RV, denoted by $\hat{M}$, with its mean value being $\bar{M}$. The randomness of $\hat{M}$ may result in the following two  cases: 1) when $\hat{M} > M$, the FRC scheme provides a higher transmission rate than the FNC scheme because there are more cooperative BSs in FRC scheme which can provide higher signal power at the typical cell-edge user, and 2) it is also possible to have $\hat{M} < M$, and in this case, the FRC scheme leads to a smaller transmission rate than the FNC scheme due to the reduction of the signal power.
When the BSs are sparse, the signal power from each BS experiences, on average, a higher path loss, and thus  the number of the cooperative BSs will significantly influence the received signal power at the cell edge users. Besides, under the condition that $\bar{M} = M$, it generally satisfies that
$\mathbb{P}\{\hat{M}\geq M\}>\mathbb{P}\{\hat{M}< M\}$
\footnote{This can be easily checked according to the PMF of the Poisson random variable.}, i.e., the number of the cooperative BSs in FRC scheme is more likely to be no less than that in FNC scheme. Therefore, the FRC scheme achieves a higher average rate when the deployment of the BSs gets sparse.
However, for dense networks, the increased  signal power brought by the more number of the cooperative BSs has a marginal effect on improving the average rate.
In this case, the difference between the average rates in FNC and FRC strategies will be dominant by the event that $\hat{M} < M$, and thus the FNC scheme achieves a higher average rate.

\subsection{Outage Probability of the Typical Cell-Edge User}
In this part, we evaluate the outage probabilities of the typical cell-edge user in BS cooperation mmWave networks.
\begin{figure}
    \centering
  \subfigure[Outage probability in FNC and FRC strategies.]{
    \label{Out1:FNCFRC} %% label for first subfigure
    \includegraphics[width=2.7 in]{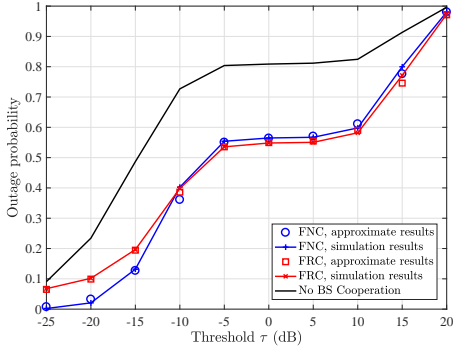}}
    \hspace{.1in}
  \subfigure[The time consumption to obtain the approximate and simulation results.]{
    \label{OutvsThreapp:sub2} %% label for second subfigure
    \includegraphics[width=2.7 in]{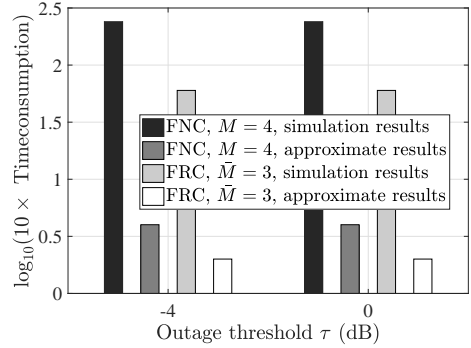}}
    \caption{Outage probability of the typical cell-edge user under FNC and FRC scheme.}
  \label{Out1}
\end{figure}

We plot the outage probabilities of the FNC and FRC strategies against the outage thresholds in Fig. \ref{Out1:FNCFRC}.
The simulation parameters are given by $M = \bar{M} = 3$, $N_L=4$, $N_N=1$, $\rho=90$ (m) and $p_L=0.2$.
The non-cooperative cases are also provided for comparison.
As the figure shows, both the FNC and FRC strategies strongly reduce the outage probabilities within a vast range of outage thresholds, which is the direct result of the increased signal power with the aid of BS cooperation.
The accuracy of the approximate method proposed in Sec. IV-B is verified in this figure. As we can see, the approximate results approach closely to the simulation results.
Fig. \ref{OutvsThreapp:sub2} plots the time consumption to calculate the simulation results and the approximation results.
Note that in Fig. \ref{OutvsThreapp:sub2}, the metric of the vertical axis is set as the logarithm of \emph{10$\times$(Time Consumption (s))}.
Obviously, the approximation method are much more computationally efficient than to directly carry out the simulation.
In Fig. \ref{Out1:FNCFRC}, we observe that when the outage threshold is small,  the outage probability in the FRC scheme becomes larger than that in the FNC scheme. This is due to the non-zero probability mass of $T^{D_{CO},\mathrm{FR}}$ at $T^{D_{CO},\mathrm{FR}}=0$, which suits to the case when there happens to have no BS in the cooperative region of the FRC scheme.
However, it presents the opposite results when the outage threshold is high. This is because the number of
the cooperative BSs in the FRC scheme may exceed that in the FNC scheme, and thus the FRC scheme may opportunistically provide higher signal power.

\begin{figure}
    \centering
  \subfigure[FNC scheme]{
    \label{Out2:FNC} %% label for first subfigure
    \includegraphics[width=2.7 in]{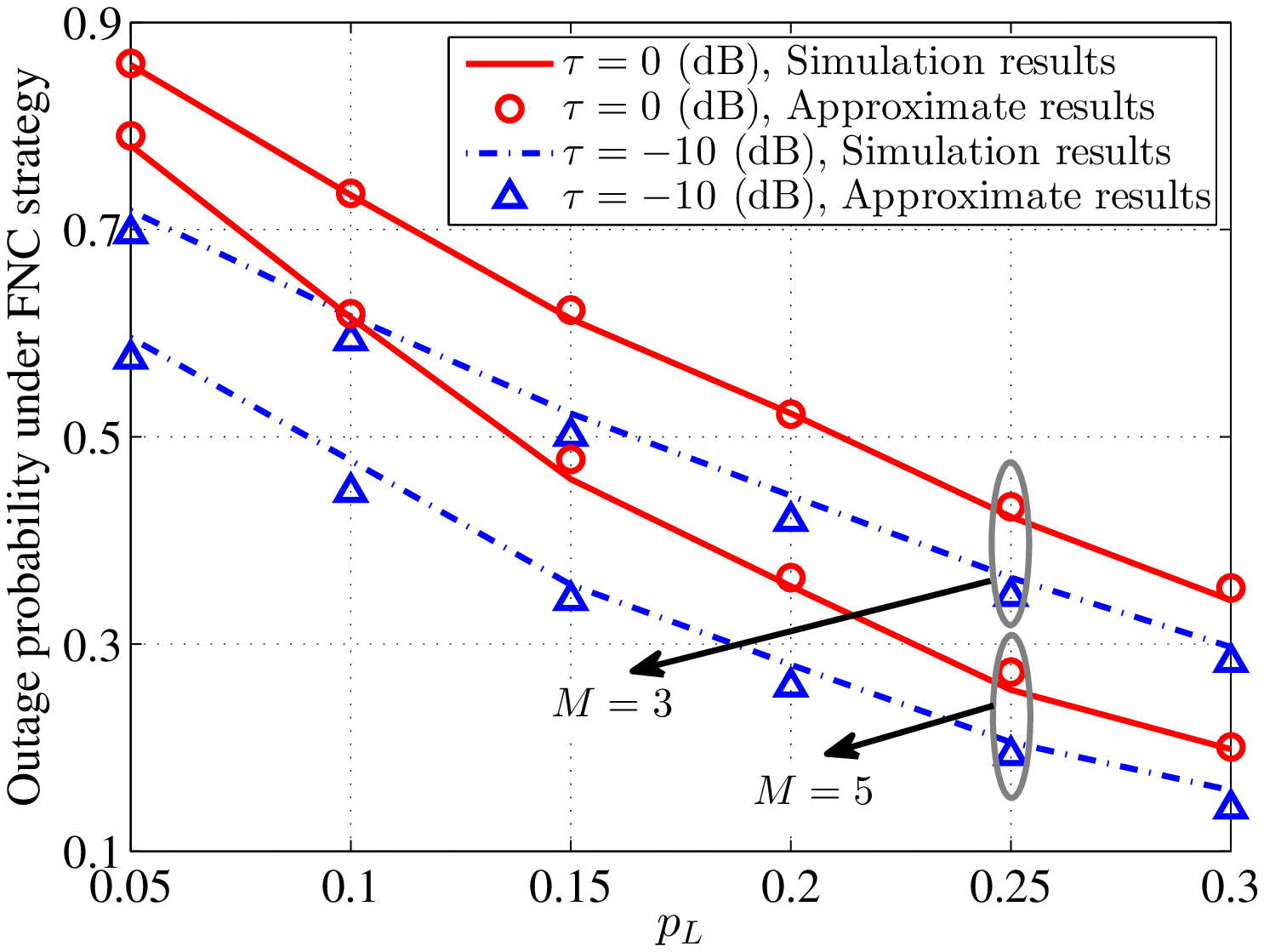}}
    \hspace{.1in}
  \subfigure[FRC scheme]{
    \label{Out2:FRC} %% label for second subfigure
    \includegraphics[width=2.7 in]{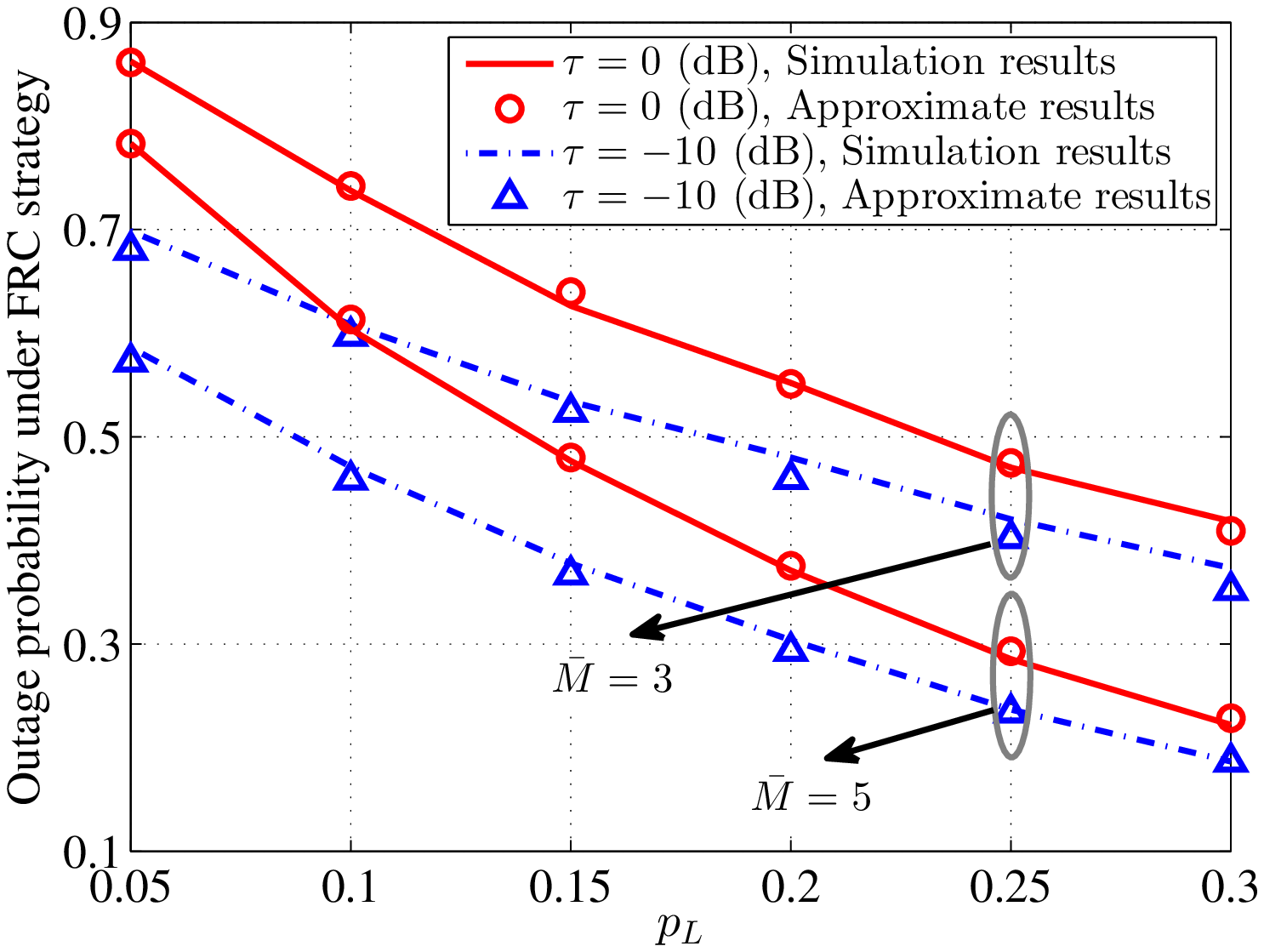}}
  \caption{Outage probability of the typical cell-edge user under FNC and FRC scheme.}
  \label{Out2} %% label for entire figure
\end{figure}

In Fig. \ref{Out2:FNC} and Fig. \ref{Out2:FRC}, we plot the outage probabilities of the FNC and FRC strategies versus the probabilities of a BS within the LOS region being in LOS state, i.e., $p_L$.
We set $N_L=4$, $N_N=1$ and $\rho=70$ (m) in the simulations and other simulation parameters are the same as those in Fig. \ref{RATE1}.
In general, a smaller value of $p_L$ indicates that the wireless signal are more likely to be blocked by the barriers and the communication links are more likely to be in NLOS state.
Fig. \ref{Out2} reveals that $p_L$ is an extremely important parameter of the BS cooperation mmWave networks.
As we can see, when $p_L$ increases from $0.05$ to $0.3$, significant decreases of the outage probabilities are observed.
This is because the LOS links provide much higher power than the NLOS links in mmWave band.

\subsection{Average Performance of a Typical General User}
\begin{figure}
    \centering
    \subfigure[Average rate.]{
    \includegraphics[width=2.7 in]{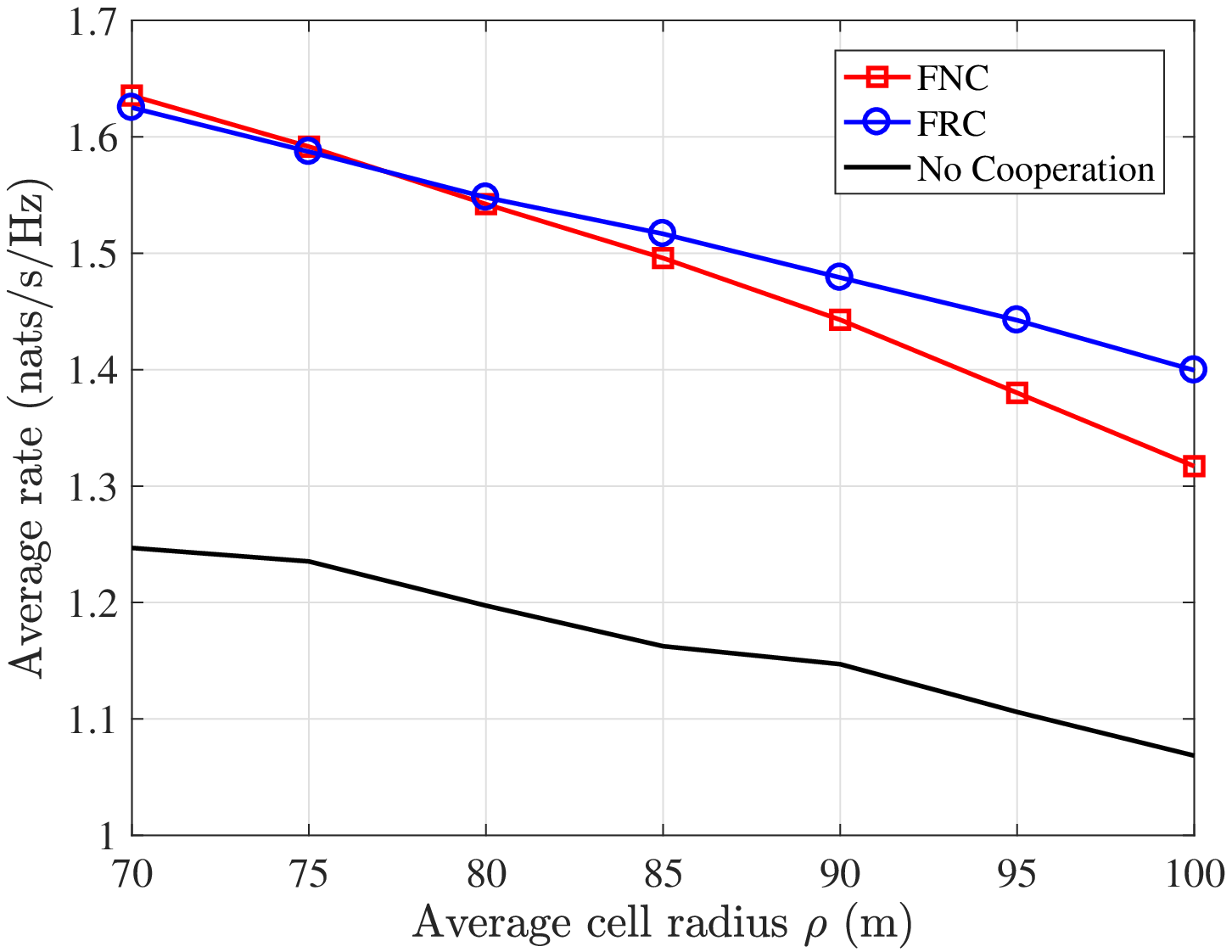}
    \label{CompNoCompA}}
    \hspace{.1in}
    \subfigure[Outage probability when the average cell radius $\rho = 90$ (m).]{
    \includegraphics[width=2.7 in]{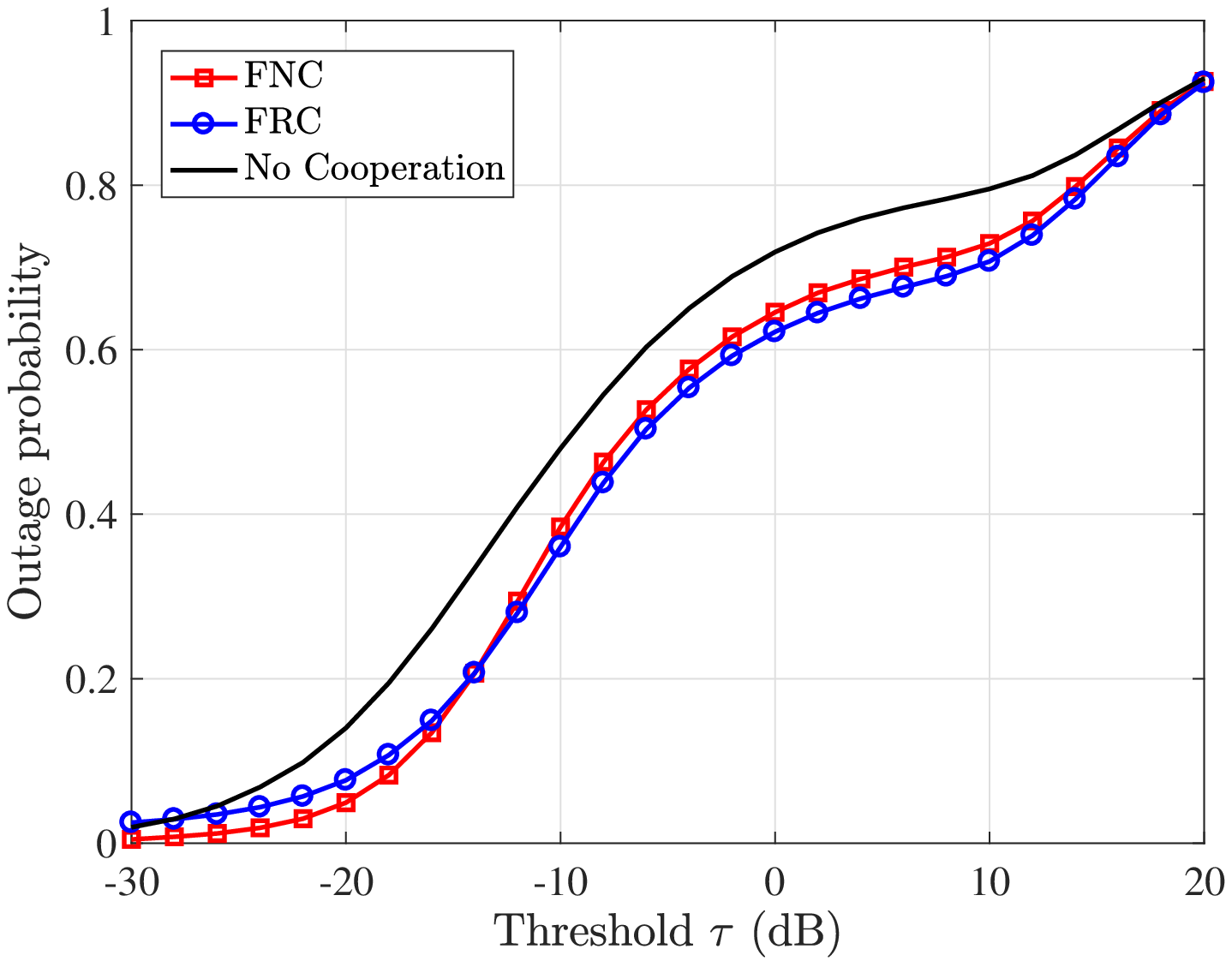}
    \label{CompNoCompB}}
    \caption{Average rate and outage probability of a typical general user.}
    \label{InPaperCoMPandNoCoMP}
    \end{figure}
Though in this paper, we focus on the performance of the cell-edge users.
It is important to evaluate the average performance of the general users, which is plotted in Fig. \ref{InPaperCoMPandNoCoMP}. The system settings are given by $M = 3$, $\bar{M} = 3$, $N_L = 4$, and $p_L = 0.2$, and other parameters are the same as those in Fig. \ref{RATE1}.
In the simulation, we assume that there is a general user located at the origin which is referred as the typical user. The BSs are spatially distributed following a HPPP.
The typical user becomes a cell-edge user if its distance to the nearest BS exceeds $D_E$, and in this case, BS cooperation will be applied to improve its performance.
If the typical user is not a cell-edge user, then it is only served by the nearest BS.
In Fig. \ref{CompNoCompA}, we plot the average rates of the typical user versus the average cell radiuses.
From Fig. \ref{CompNoCompA}, we can see that in all cases, the average rate can be significantly improved with the aided of BS cooperation. For example, when $\rho = 100$ (m), the FNC and FRC strategies can increase the average rates by around $24\%$ and $32\%$, respectively.
Fig. \ref{CompNoCompB} illustrates the outage probability of the typical user. In Fig. \ref{CompNoCompB},
within a wide range of the outage threshold, significant reduction of the outage probability can be observed with the help of the FNC and FRC strategies.
From the simulation results in Fig. \ref{InPaperCoMPandNoCoMP}, we conclude that if the BS cooperation is applied to cell edge users, the average performance of the general users can be greatly improved.

\section{Conclusion}
\label{Section6}
In this paper, the performance of the edge users in BS cooperation-aided mmWave cellular networks was detailedly investigated based on the stochastic geometry framework.
The expressions for the average rate and outage probability for a typical cell-edge user under two different cooperative strategies, i.e., FNC and FRC, were analytically  derived.
For the outage probability, we also propose to approximate the signal power by using Gamma RVs. The approximate results are computationally efficient than the analytical expressions.
Numerical results show that the derived expressions are very accurate and time-saving. Simulations also demonstrate that the lack of LOS and the severe path loss of NLOS have a great impact on the performance of the edge users, which will be significantly improved by applying the BS cooperation techniques to the mmWave cellular networks.

\appendices
\section{Proof of Theorem 2 and Theorem 3}
\label{Appendix:LaplaceTransInFNC}
\subsection{The derivation of $\mathcal{L}_{T_k^{M,\mathrm{FN}}|r}\left(z\right)$}
We first have
\begin{align}
\label{DerivationLTSubT}
\mathcal{L}_{T_{k}^{M,\mathrm{FN}}|r}\left(z\right)=
\mathbb{E}_{D^{(k)},L_{(k)},h_{(k)}}\left\{
\exp\left({-zG_MP_Th_{(k)}L_{(k)}\left(D^{(k)}\right)}\right)\bigg|D_{M+1}=r\right\},
\end{align}
where the conditional PDF of $D^{(k)}$ is given in \eqref{DistanceConditionPDF}.
When $r\leq D$, the $k^{\mathrm{th}}$ link is in LOS (NLOS) state with probability $p_L$ ($p_N$).
When $r > D$, if $D^{(k)}\leq D$, then the $k^{\mathrm{th}}$ link is in LOS (NLOS) state with probability $p_L$ ($p_N$), otherwise, it is always in NLOS state. Therefore, for $\nu\in\{L,N\}$, we have
\begin{align}
\mathcal{L}_{T_{k}^{M,\mathrm{FN}}|r}\left(z\right) & =\left\{
\begin{aligned}
&\sum_{\nu\in\{L,N\}}\int_{D_E}^{r}
\frac{2p_{\nu}y\mathcal{L}_{h_{(k)}^{(\nu)}}
\left(za_{M,{\nu}}y^{-\alpha_{\nu}}\right)\mathrm{d}y}{r^2-D_E^2},&&r\leq D\\
&\sum_{\nu\in\{L,N\}}\int_{D_E}^{r}
\frac{2p_{\nu}y\mathcal{L}_{h_{(k)}^{(\nu)}}\left(za_{M,{\nu}}y^{-\alpha_{\nu}}\right)\mathrm{d}y}{r^2-D_E^2}+ \int_{D}^{r}\frac{2y\mathcal{L}_{h_{(k)}^{(N)}}\left(za_{M,N}y^{-\alpha_{N}}\right)\mathrm{d}y}{r^2-D_E^2},&&r> D,
\end{aligned}\right.
\nonumber \\
& =\left\{
\begin{aligned}
&\frac{\sum_{\nu\in\{L,N\}} p_{\nu}\Lambda_{M,\nu}\left(z,r,D_E\right)}{r^2-D_E^2},&&r\leq D\\
&\frac{
\sum_{\nu\in\{L,N\}} p_{\nu}\Lambda_{M,\nu}\left(z,D,D_E\right)
+\Lambda_{M,N}\left(z,r,D\right)
}{r^2-D_E^2},&&r> D.
\end{aligned}\right.
\end{align}

\subsection{The derivation of $\mathcal{L}_{I_1^{M,\mathrm{FN}}|r}(z)$}
If $r \leq D$, then the interfering link between the $(M+1)^{\mathrm{th}}$ nearest BS and the typical edge user may be in LOS or NLOS state. If $r > D$, then the interfering link is always in NLOS state. Therefore, we have
\begin{align}
\mathcal{L}_{I_1^{M,\mathrm{FN}}|r}\left(z\right)=\left\{
\begin{aligned}
&\mathbb{E}_{\theta_{M+1}}\left[
    \mathbb{E}_{h_{M+1},L_{M+1}}\left[
    \exp\left[{-zP_TG_B\left(\theta_{M+1}\right)L_{M+1}\left(D_{M+1}\right)}\right]
    \right]
    \Big|D_{M+1}=r\right],&r\leq D\\
&\mathbb{E}_{\theta_{M+1}}\left[\mathbb{E}_{h_{M+1}^{(N)}}
    \left[
    \exp\left[{-zP_TG_B\left(\theta_{M+1}\right)h_{M+1}^{(N)}C_ND_{M+1}^{-\alpha_N}}\right]
    \right]
    \Big|D_{M+1}=r\right],&r>D
\end{aligned}\right.\nonumber
\end{align}
Calculating the expectation operations directly leads to \eqref{At_rInterferenceLaplaceTransFNC}.

\subsection{The derivation of  $\mathcal{L}_{I_2^{M,\mathrm{FN}}|r}(z)$}
To obtain $\mathcal{L}_{I_2^{M,\mathrm{FN}}|r}\left(z\right)$, there are two different cases, i.e., (1) $r \leq D$, and (2) $r > D$. Due to space limitation, we only provide the proof for the case when $r\leq D$. The same procedure can be extended to obtain the results when $r>D$.
If $r \leq D$, then the interfering BSs in $\mathcal{R}^2\backslash\mathcal{B}\left(o,r\right)$ can be classified into six independent PPPs defined in Table I, and therefore, we have
    \begin{align}
      \mathcal{L}_{I_2^{M,\mathrm{FN}}|r}\left(z\right)=
      \left(\prod_{\mu\in\{M,S\}}\mathcal{L}_{I_{\mu,L}|r}\left(z\right)\right)
      \left(\prod_{\mu\in\{M,S\}}\mathcal{L}_{I_{\mu,N}^{\leq D}|r}\left(z\right)\right)
      \left(\prod_{\mu\in\{M,S\}}\mathcal{L}_{I_{\mu,N}^{>D}|r}\left(z\right)\right)
      \label{prodLT}
    \end{align}
    where $I_{\mu,L}$, $I_{\mu,N}^{\leq D}$ and $I_{\mu,N}^{>D}$ ($\mu\in\{M,S\}$) are the summation of interferences from the BSs in $\Phi_{\mu,L}$, $\Phi_{\mu,N}^{\leq D}$ and $\Phi_{\mu,N}^{>D}$, respectively, defined in Table I. In \eqref{prodLT}, we further have
    \begin{subequations}
    \begin{align}
      \mathcal{L}_{I_{\mu,L}|r}\left(z\right) &=
     \exp\left( -2\pi p_{\mu}p_L\lambda_B \int_r^D\left(1-\mathcal{L}_{h^{(L)}}\left(z a_{\mu,L}y^{-\alpha_{L}}\right)\right) y\mathrm{d}y\right),
      \label{LTInterferenceNuL}\\
      \mathcal{L}_{I_{\mu,N}^{\leq D}|r}\left(z\right)&= \exp\left( -2\pi p_{\mu}p_N\lambda_B \int_r^D\left(1-\mathcal{L}_{h^{(N)}}\left(z a_{\mu,N}y^{-\alpha_{N}}\right)\right) y\mathrm{d}y\right),
      \label{LTInterferenceNuNwithinD}\\
      \mathcal{L}_{I_{\mu,N}^{>D}|r}\left(z\right) &= \exp\left( -2\pi p_{\mu}\lambda_B \int_D^{+\infty}\left(1-\mathcal{L}_{h^{(N)}}\left(z a_{\mu,N}y^{-\alpha_{N}}\right)\right) y\mathrm{d}y\right),
      \label{LTInterferenceNuNoutsideD}
    \end{align}
    \label{LTin}
    \end{subequations}
    where the three equations above follow from the PGFL of PPP \cite{Book:SG2013}.
    \begin{table}[t]
    \label{InterferenceClassSmall}
    \centering
    \caption{Classification of the Interferences when $r\leq D$}
     \begin{tabular}{|c||c|c|}
     \hline
     \hline
        PPP & Properties & Density\\
     \hline
        $\Phi_{M,L}$&  LOS, main lobe interference; Located in $\mathcal{R}\left(r,D\right)$  &$p_Mp_L\lambda_B$\\
     \hline
        $\Phi_{S,L}$& LOS, side lobe interference; Located in $\mathcal{R}\left(r,D\right)$;  &$p_Sp_L\lambda_B$\\
     \hline
        $\Phi_{M,N}^{(\leq D)}$ &  NLOS, main lobe interference; Located in $\mathcal{R}\left(r,D\right)$ &$p_Mp_N\lambda_B$  \\
     \hline
        $\Phi_{S,N}^{(\leq D)}$ &  NLOS, side lobe interference; Located in $\mathcal{R}\left(r,D\right)$ &$p_Sp_N\lambda_B$  \\
     \hline
        $\Phi_{M,N}^{(> D)}$ &  NLOS, main lobe interference; Located in $\mathcal{B}^{c}\left(o,D\right)$ &$p_M\lambda_B$  \\
     \hline
        $\Phi_{S,N}^{(> D)}$ &  NLOS, side lobe interference; Located in $\mathcal{B}^{c}\left(o,D\right)$ &$p_S\lambda_B$  \\
     \hline
     \hline
     \end{tabular}
    \end{table}
Substituting \eqref{LTInterferenceNuL}, \eqref{LTInterferenceNuNwithinD}, and \eqref{LTInterferenceNuNoutsideD} into \eqref{prodLT}, we obtain the result in \eqref{Outof_rInterferenceLaplaceTransFNC} for $r\leq D$.

\section{Proof of Theorem 5}\label{Appendix:LaplaceTransInFRC}
\subsection{The derivation of $\mathcal{L}_{T^{D_{CO},\mathrm{FR}}}(z)$}
Following the PGFL of PPP \cite{Book:SG2013}, we have
\begin{align}
\mathcal{L}_{T^{D_{CO},\mathrm{FR}}}\left(z\right)
&=\exp\left(-2\pi\lambda_B\int_{D_E}^{D_{CO}}
\left[1-\mathbb{E}_{L,h}\left(\exp\left(-zP_TG_MhL\left(y\right)\right)\right)
\right]y\mathrm{d}y\right) \nonumber\\
&=\exp\left(-\pi\lambda_B\left(D_{CO}^2-D_E^2\right)\right)
\exp\left(2\pi\lambda_B\int_{D_E}^{D_{CO}}\mathbb{E}_{L,h}
\left(\exp\left(-zP_TG_MhL\left(y\right)\right)\right)y\mathrm{d}y\right).
\nonumber
\end{align}
Calculating the integral in the exponential part directly leads to \eqref{FRCSignalLaplaceTrans}.

\subsection{The derivation of $\mathcal{L}_{I^{D_{CO},\mathrm{FR}}}(z)$}
In fact, $I^{D_{CO},\mathrm{FR}}$ is the total interferences caused by the BSs located in $\mathcal{R}^2\backslash \mathcal{B}\left(o,D_{CO}\right)$. According to the definition of $I_2^{M,\mathrm{FN}}$ given in Theorem 3, if we have $D_{M+1}=D_{CO}$, then $I^{D_{CO},\mathrm{FR}}$ and $I_2^{M,\mathrm{FN}}$ are two identical distributed RVs. Thus, we have \eqref{FRCInterferenceLaplaceTrans}.

\section{The calculation of $\kappa_\nu^{M,\mathrm{FN}}$, $\rho_\nu^{M,\mathrm{FN}}$, $\kappa_\nu^{D_{CO},\mathrm{FR}}$ and $\rho_\nu^{D_{CO},\mathrm{FR}}$ for $\nu\in\{L,N\}$}
\label{APP:3}
\subsection{$\kappa_L^{M,\mathrm{FN}}$ and $\rho_L^{M,\mathrm{FN}}$}
According to \eqref{GammaApproximate1}, we need the first and second order moments of $\breve{T}_L^{M,\mathrm{FN}}$.
The first moment of $\breve{T}_L^{M,\mathrm{FN}}$ is calculated as
\begin{align}
&\mathbb{E}\left(\breve{T}_L^{M,\mathrm{FN}}\bigg|S_L^{M,\mathrm{FN}}\geq1\right)= \mathbb{E}_{D_{M+1}}\left\{\mathbb{E}_{h_{(i)}^{(L)},D^{(i)},S_L^{M,\mathrm{FN}}|r}
\left(\sum_{ i \in \Omega_{L}^{M,\mathrm{FN}}} a_{M,L}h_{(i)}^{(L)}\left(D^{(i)}\right)^{-\alpha_L}\bigg|S_L^{M,\mathrm{FN}}\geq1\right)\right\}\nonumber \\
&=\int_{D_E}^{+\infty}f_{M+1}\left(r\right)\Bigg\{
\mathbb{E}_{S_L^{M,\mathrm{FN}}|r}\left(S_L^{M,\mathrm{FN}}\bigg|S_L^{M,\mathrm{FN}}\geq1\right)
\mathbb{E}_{h_{(i)}^{(L)},D^{(i)}|r}\left(a_{M,L}h_{(i)}^{(L)}\left(D^{(i)}\right)^{-\alpha_L}\right)
\Bigg\}\mathrm{d}r,
\label{FirstOrderFNCL}
\end{align}
where we have $\mathbb{E}_{X|r}(\cdot)\triangleq\mathbb{E}_{X}(\cdot|D_{M+1}=r)$ and
\begin{align}
\mathbb{E}_{S_L^{M,\mathrm{FN}}|r}\left(S_L^{M,\mathrm{FN}}\bigg|S_L^{M,\mathrm{FN}}\geq1\right)& = \sum_{m=1}^M m\mathcal{P}_{S_L^{M,\mathrm{FN}}}\left(m,r\right), \\
\mathbb{E}_{h_{(i)}^{(L)},D^{(i)}|r}\left(a_{M,L}h_{(i)}^{(L)}\left(D^{(i)}\right)^{-\alpha_L}\right)
&=\left\{
\begin{aligned}
&\frac{2a_{M,L}\left(\mathrm{min}\left(r,D\right)^{2-\alpha_L}-D_E^{2-\alpha_L}\right)}
{\left(2-\alpha_L\right)\left(\mathrm{min}\left(r,D\right)^{2}-D_E^{2}\right)},&\alpha_L>2\\
&\frac{2a_{M,L}\left[\ln\left(\mathrm{min}\left(r,D\right)\right)-\ln\left(D_E\right)\right]}
{\left(\mathrm{min}\left(r,D\right)^{2}-D_E^{2}\right)},&\alpha_L=2
\end{aligned}\right.
\end{align}
with $\mathcal{P}_{S_L^{M,\mathrm{FN}}}\left(m,r\right)$ defined in \eqref{PMFNumLOS}.
The second order moment is calculated as
\begin{align}
&\mathbb{E}\left(\left(\tilde{T}_L^{M,\mathrm{FN}}\right)^2\bigg|S_L^{M,\mathrm{FN}}\geq1\right) \nonumber\\
&= \int_{D_E}^{+\infty}f_{M+1}\left(r\right)\Bigg\{
\mathbb{E}_{S_L^{M,\mathrm{FN}}|r}\left[S_L^{M,\mathrm{FN}}\bigg|S_L^{M,\mathrm{FN}}\geq1\right]
\mathbb{E}_{h_{(i)}^{(L)},D^{(i)}|r}\left[\left(a_{M,L}h_{(i)}^{(L)}\left(D^{(i)}\right)^{-\alpha_L}\right)^2\right]
\nonumber \\
&\quad+
\mathbb{E}_{S_L^{M,\mathrm{FN}}|r}\left[S_L^{M,\mathrm{FN}}\left(S_L^{M,\mathrm{FN}}-1\right)\bigg|S_L^{M,\mathrm{FN}}\geq1
\right]
\left[\mathbb{E}_{h_{(i)}^{(L)},D^{(i)}|r}\left(a_{M,L}h_{(i)}^{(L)}
\left(D^{(i)}\right)^{-\alpha_L}\right)\right]^2
\Bigg\}\mathrm{d} r
\label{SecondOrderFNCL}
\end{align}
where we have
\begin{align}
&\mathbb{E}_{S_L^{M,\mathrm{FN}}|r}\left[S_L^{M,\mathrm{FN}}\left(S_L^{M,\mathrm{FN}}-1\right)\bigg|S_L^{M,\mathrm{FN}}\geq1\right] = \sum_{m=1}^M m\left(m-1\right)\mathcal{P}_{S_L^{M,\mathrm{FN}}}\left(m,r\right),\\
&\mathbb{E}_{h_{(i)}^{(L)},D^{(i)}|r}\left[\left(a_{M,L}h_{(i)}^{(L)}\left(D^{(i)}\right)^{-\alpha_L}\right)^2\right] =
\frac{\left(N_L+1\right)a_{M,L}^2\left(\mathrm{min}\left(r,D\right)^{2-2\alpha_L}-D_E^{2-2\alpha_L}\right)}{N_L\left(1-\alpha_L\right)
\left(\mathrm{min}\left(r,D\right)^2-D_E^2\right)}.
\end{align}
Inserting \eqref{FirstOrderFNCL} and \eqref{SecondOrderFNCL} into \eqref{GammaApproximate1}, we can obtain $\kappa_L^{M,\mathrm{FN}}$ and $\rho_L^{M,\mathrm{FN}}$.

\subsection{$\kappa_N^{M,\mathrm{FN}}$ and $\rho_N^{M,\mathrm{FN}}$}
According to \eqref{GammaApproximate2}, to obtain $\kappa_N^{M,\mathrm{FN}}$ and $\rho_N^{M,\mathrm{FN}}$, we need the first and second order moments of
$T_N^{M,\mathrm{FN}}$. In fact, the calculation of $\mathbb{E}\left[T_N^{M,\mathrm{FN}}\right]$ and $\mathbb{E}\left[\left(T_N^{M,\mathrm{FN}}\right)^2\right]$ are similar to
\eqref{FirstOrderFNCL} and \eqref{SecondOrderFNCL}. The only difference is that in \eqref{FirstOrderFNCL} and \eqref{SecondOrderFNCL}, $S_L^{M,\mathrm{FN}}$ is a RV, while the cardinality of $\Omega_{N}^{M,\mathrm{FN}}$, is no longer a RV but a fixed constant, i.e., $M$.

\subsection{$\kappa_\nu^{D_{CO},\mathrm{FR}}$ and $\rho_\nu^{D_{CO},\mathrm{FR}}$ for $\nu\in\{L,N\}$}
Due to the space limitation, we only provide the result for $\kappa_L^{D_{CO},\mathrm{FR}}$ and $\rho_L^{D_{CO},\mathrm{FR}}$ when $D_{CO}\leq D$.
Following similar steps, we can obtain the results for other cases.
Note that the cooperative BSs in LOS state are distributed as a PPP within $\mathcal{R}\left(o,D_E,D_{CO}\right)$ with density $\lambda_L=p_L\lambda_B$.
Therefore, conditioning on there is at least one BS in LOS state, we have
\begin{align}
&\mathbb{E}\left(\tilde{T}_{L}^{D_{CO},\mathrm{FR}}\right)= \frac{2\pi a_{M,L}\lambda_L}{1-p_{0,L}^{D_{CO},\mathrm{FR}}}\int_{D_E}^{D_{CO}}  r^{1-\alpha_L}\mathrm{d}r=\left\{
\begin{aligned}
&\frac{2\pi a_{M,L}\lambda_L\left[D_{CO}^{2-\alpha_L} - D_E^{2-\alpha_L}\right]}{\left(1-p_{0,L}^{D_{CO},\mathrm{FR}}\right)\left(2-\alpha_L\right)}, &\alpha_L>2,\\
&\frac{2\pi a_{M,L}\lambda_L\left[\ln\left(D_{CO}\right)  - \ln\left(D_E\right)\right]}{\left(1-p_{0,L}^{D_{CO},\mathrm{FR}}\right)}, &\alpha_L=2,
\end{aligned}\right.
\nonumber \\
&\mathbb{E}\left(\left(\tilde{T}_{L}^{D_{CO},\mathrm{FR}}\right)^2\right) =
\frac{\frac{N_L+1}{N_L}\int_{D_E}^{D_{CO}}2\pi a_{M,L}^2\lambda_Lr^{1-2\alpha_L}\mathrm{d}r
+\left(\int_{D_E}^{D_{CO}}2\pi a_{M,L}\lambda_Lr^{1-\alpha_L}\mathrm{d}r\right)^2}{1-p_{0,L}^{D_{CO},\mathrm{FR}}} \nonumber \\
&=\frac{1}{1-p_{0,L}^{D_{CO},\mathrm{FR}}}
\Bigg\{\frac{\lambda_L\pi a_{M,L}^2\left(N_L+1\right)}{N_L\left(1-\alpha_L\right)}\left(D_{CO}^{2-2\alpha_L}-D_E^{2-2\alpha_L}\right) \nonumber \\
&+\left(2\pi a_{M,L}\lambda_L\right)^2\left[\mathbb{I}\left(\alpha_L>2\right)\frac{\left(D_{CO}^{2-\alpha_L}-D_E^{2-\alpha_L}\right)}{2 - \alpha_L}
+\mathbb{I}\left(\alpha_L=2\right)\left(\ln\left(D_{CO}\right)-\ln\left(D_E\right)\right)\right]^2\Bigg\}
\nonumber
\end{align}
Inserting these results into \eqref{GammaApproximate3}, we obtain $\kappa_L^{D_{CO},\mathrm{FR}}$ and $\rho_L^{D_{CO},\mathrm{FR}}$.


\begin{thebibliography}{99}
\bibitem{What5GBe}
J. G. Andrew \emph{et al.}, ``What will 5G be?'' \emph{IEEE J. Sel. Areas Commun.,} vol. 32, no. 6, pp. 1065--1082, Jun. 2014.



\bibitem{A.Pi2011}
A. Pi and F. Khan, ``An introduction to millimeter--wave mobile broadband systems,'' \emph{IEEE Commun. Mag.,} vol. 49, no. 6, pp. 101--107, Jun. 2011.

\bibitem{FiveDisruptive}
F. Boccardi, R. W. Heath, A. Lozano, T. L. Marzetta, and P. Popovski, ``Five disruptive technology directions for 5G,'' \emph{IEEE Commun. Mag.}, vol. 52, no. 2,pp. 74--80, Feb. 2014.

\bibitem{T.Pappaport2013}
T. Pappaport \emph{et al.}, ``Millimeter wave mobile communications for 5G cellular: It will work!'' \emph{IEEE Access}, vol. 1, pp. 335--349, May 2013.


\bibitem{S.Rangan2014}
S. Rangan, T.S. Rappaport and E. Erkip,``Millimeter--wave cellular wireless networks: Potentials and challenges,'' \emph{Proc. IEEE, } vol. 102, no. 2, pp. 366--385, Mar. 2014.



\bibitem{M.R.Akdeniz2014}
M. R. Akdeniz, Y. Liu, M. K. Samimi, S. Sun, S. Rangan, T. S.
Rappaport, and E. Erkip, ``Millimeter wave channel modeling and cellular
capacity evaluation,'' \emph{IEEE J. Sel. Areas Comm.}, vol. 32, no. 6, pp.
1164--1179, Jun. 2014.

\bibitem{T.Bai2015}
T. Bai and R. Heath, ``Coverage and rate analysis for millimeter-wave cellular networks''
\emph{IEEE Trans. Wireless Commun.} vol. 14, no. 2, pp. 1100--1114, Feb. 2015.



\bibitem{V.Petrov2017}
V. Petrov, M. Komarov, D. Moltchanov, J. M. Jornet, and Y. Koucheryavy, ``Interference and SINR in millimeter wave and terahertz communication systems with blocking and directional antennas,'' \emph{IEEE Trans. Wireless Commun.,} vol. 16, no. 3, pp. 1791--1808, March 2017.


\bibitem{M.Sawahashi2010}
M. Sawahashi, Y. Kishiyama, A. Morimoto, D. Nishikawa, and
M. Tanno, ``Coordinated multipoint transmission/reception techniques
for LTE-Advanced [coordinated and distributed MIMO],'' \emph{IEEE Wireless
Commun. Mag.,} vol. 17, no. 3, pp. 26--34, Jun. 2010.

%\bibitem{}
%Coordinated Multipoint Joint Transmission in Heterogeneous Networks

\bibitem{D.Lee2012}
D. Lee et al., ``Coordinated multipoint transmission and reception in
LTE-Advanced: Deployment scenarios and operational challenges,'' \emph{IEEE
Commun. Mag.,} vol. 50, no. 2, pp. 148--155, Feb. 2012.

\bibitem{X.Tao2012}
X. Tao, X. Xu, and Q. Cui, ``An overview of cooperative communications,'' \emph{IEEE Commun. Mag.,} vol. 50, no. 6, pp. 65--71, Jun. 2012.

\bibitem{R.Irmer2011}
R. Irmer \emph{et al}., ``Coordinated multipoint: Concepts, performance, field
trial results,'' \emph{IEEE Commun. Mag.,} vol. 49, no. 2, pp. 102--111, Feb. 2011.

%ºÁÃײ¨²¿·Ö
\bibitem{T.Bai2014}
T. Bai, A. Alkhateeb, and R. Heath, ``Coverage and capacity of millimeter-wave cellular networks,'' \emph{IEEE Commun. Mag.,} vol. 52, no. 9, pp. 70--77, Sep. 2014.


\bibitem{S.Singh2015}
S. Singh, M. Kulkarni, A. Ghosh and J. Andrews, ``Tractable model for rate in self--backhauled millimeter wave cellular networks,'' \emph{IEEE J. Sel. Areas Commun.,} vol. 33, no. 10, pp. 2196--2211, Oct. 2015.




\bibitem{M.Renzo2015}
M. D. Renzo, ``Stochastic geometry modeling and analysis of multi-tier millimeter wave cellular networks'' \emph{IEEE Trans. Wireless Commun.}, vol. 14, no. 9, pp. 5038--5057, Sep. 2015.



{
\bibitem{E.Turgut2017TCOM}
E. Turgut and M. C. Gursoy, ``Coverage in heterogeneous downlink millimeter wave cellular networks,'' \emph{IEEE Trans. Commun.}, vol. 65, no. 10, pp. 4463--4477, Oct. 2017.
}

\bibitem{A.Thornburg2016}
A. Thornburg, T. Bai and R. Heath, ``Performance analysis of outdoor mmWave Ad hoc networks,'' \emph{IEEE Trans. signal process.,} vol. 64, no. 15, pp. 4065--4079, Aug. 2016.

{
\bibitem{C.Wang2016TWC}
C. Wang and H.-M. Wang, ``Physical layer security in millimeter wave cellular networks,'' \emph{IEEE Trans. Wireless Commun.,} vol. 15. no. 8, pp. 5569--5585, Aug. 2016.
}
%SG

\bibitem{M.Haenggi2009}
M. Haenggi, J. G. Andrews, F. Baccelli, O. Dousse, and M.
Franceschetti, ``Stochastic geometry and random graphs for the analysis
and design of wireless networks,'' \emph{IEEE J. Sel. Areas Commun.}, vol.
27, no. 7, pp. 1029--1046, Sep. 2009.



\bibitem{J.Andrews2011}
J. Andrews, F. Baccelli, and R. Ganti, ``A tractable approach to coverage and rate in cellular networks,'' \emph{IEEE Trans. Commun.}, vol.59, no.11, pp. 3122--3134, Nov. 2011.


\bibitem{P.Madhusudhanan2012}
P. Madhusudhanan, J. G. Restrepo, Y. Liu and T. X. Brown, ``Downlink coverage analysis in a heterogeneous cellular network,'' \emph{2012 IEEE Global Communications Conference (GLOBECOM)}, Anaheim, CA, 2012, pp. 4170--4175.


\bibitem{P.Madhusudhanan2016}
P. Madhusudhanan, J. G. Restrepo, Y. Liu, and T. X Brown,
``Analysis of downlink connectivity models in a heterogeneous cellular network via stochastic geometry,'' \emph{IEEE Trans. Wireless Commun.}, vol.15, no. 6, pp. 3895--3907, Jun. 2016.


\bibitem{T.-X.Zheng2017TCOM}
T.-X. Zheng, H.-M. Wang, and M. H. Lee, ``Multi-antenna transmission in downlink heterogeneous cellular networks under a threshold-based mobile association policy,'' \emph{IEEE Trans. Commun.}, vol. 65, no. 1, pp. 244--256, Jan. 2017.




%CoMP

\bibitem{G.Nigam2014}
G. Nigam, P. Minero, and M. Haenggi, ``Coordinated multipoint joint transmission in heterogeneous networks,'' \emph{IEEE Trans. Commun.,} vol. 62, no. 11, pp. 4134--4146, Nov. 2014.

\bibitem{G.Nigam2015}
G. Nigam, P. Minero and M. Haenggi, ``Spatiotemporal cooperation in heterogeneous cellular networks,'' \emph{IEEE J. Sel. Areas Commun.,} vol. 33, no. 6, pp. 1253-1265, Jun. 2015.


\bibitem{G.Nigam2015GLOBECOM}
G. Nigam and P. Minero, ``Spatiotemporal base station cooperation in a cellular network: The worst-case user,'' \emph{2015 IEEE Global Communications Conference (GLOBECOM)}, San Diego, CA, 2015.




\bibitem{Q.Cui2018TCOM}
Q. Cui, X. Yu, Y. Wang and M. Haenggi, ``The SIR Meta distribution in poisson cellular networks with base station cooperation,'' \emph{IEEE Trans. Commun.}, to be published.

\bibitem{X.Yu2017WCL}
X. Yu, Q. Cui and M. Haenggi, ``Coherent joint transmission in downlink heterogeneous cellular networks,''
\emph{IEEE Wireless Commun. Let.}, to be published.






\bibitem{R.Tanbourgi2014_1}
R. Tanbourgi, S. Singh, J. G. Andrews and F. K. Jondral, ``Analysis of non--coherent joint--transmission cooperation in heterogeneous cellular networks,'' \emph{2014 IEEE International Conference on Communications (ICC)}, Sydney, NSW, 2014, pp. 5160--5165.


\bibitem{R.Tanbourgi2014_2}
R. Tanbourgi, S. Singh, J. G. Andrews, and F. K. Jondral, ``A tractable
model for non--coherent joint--transmission base station cooperation,''
\emph{
IEEE Trans. Wireless Commun.}, vol. 13, no. 9, pp. 4959--4973, Sep.
2014.


\bibitem{W.Nie2016}
W. Nie, F. Zheng, X.-C. Wang, W. Zhang, and S. Jin,
``
User--centric cross--tier base station clustering and cooperation in heterogeneous networks: Rate improvement and energy saving''
\emph{IEEE J. Sel. Areas Commun.}, vol. 34, no. 5, pp. 1192--1206, May 2016.



\bibitem{D.Maamari2016}
D. Maamari, N. Devroye, and D. Tuninetti, ``Coverage in mmWave celluar networks with base station co--operation,'' \emph{IEEE Trans. Wireless Commun.}, vol. 15, no. 4, pp. 2981--2994, Apr. 2016.

\bibitem{R.W.Heath2013}
R. W. Heath, M. Kountouris, and T. Bai, ``Modeling heterogeneous network interference using poisson point processes,'' \emph{IEEE Trans. Signal Process.}, vol. 61, no. 16, pp. 4114--4126, Aug. 2013.

\bibitem{N.Deng2014}
N. Deng, W. Zhou, and M. Haenggi, ``A heterogeneous cellular network model with inter-tier dependence,'' \emph{2014 IEEE Global Communications Conference (GLOBECOM)}, Austin, TX, 2014, pp. 1522--1527.

\bibitem{C.h.Lee2012}
C. h. Lee and M. Haenggi, ``Interference and outage in poisson cognitive networks,'' \emph{IEEE Trans. Wireless Commun.}, vol. 11, no. 4, pp. 1392--1401, Apr. 2012.

\bibitem{Z.Yazdanshenasan2016NovTWC}
Z. Yazdanshenasan, H. S. Dhillon, M. Afshang, and P. H. J. Chong, ``Poisson hole process: Theory and applications to wireless networks,'' \emph{IEEE Trans. Wireless Commun..}, vol. 15, no. 11, pp. 7531--7546, Nov. 2016.


\bibitem{S.Han2015CM}
S. Han, C. l. I, Z. Xu, and C. Rowell, ``Large-scale antenna systems with hybrid analog and digital beamforming for millimeter wave 5G,'' \emph{IEEE Commun. Mag.}, vol. 53, no. 1, pp. 186--194, Jan. 2015.



\bibitem{O.E.Ayach2014TWC}
O. E. Ayach, S. Rajagopal, S. Abu-Surra, Z. Pi, and R. W. Heath, ``Spatially sparse precoding in millimeter wave MIMO systems,'' \emph{IEEE Trans. Wireless Commun.}, vol. 13, no. 3, pp. 1499--1513, Mar. 2014.



\bibitem{C.Jeong2015CM}
C. Jeong, J. Park, and H. Yu, ``Random access in millimeter-wave beamforming cellular networks: Issues and approaches,'' \emph{IEEE Commun. Mag.}, vol. 53, no. 1, pp. 180--185, Jan. 2015.



\bibitem{Y.Li2017TWC}
Y. Li, J. G. Andrews, F. Baccelli, T. D. Novlan, and C. J. Zhang, ``Design and analysis of initial access in millimeter wave cellular networks,'' \emph{IEEE Trans. Wireless Commun.}, vol. 16, no. 10, pp. 6409--6425, Oct. 2017.




\bibitem{K.A.Hamdi2008}
K. A. Hamdi, ``Capacity of MRC on correlated rician fading channels,'' in \emph{IEEE Trans. Commun.}, vol. 56, no. 5, pp. 708--711, May 2008.


\bibitem{J.Gil-Pelaez1951}
J. Gil-Pelaez, ``Note on the inversion theorem, '' \emph{Biometrika}, vol. 49, no. 2, pp. 481--482, 1951.



\bibitem{Book:ISG2007}
I. S. Gradshteyn, I. M. Ryzhik, A. Jeffrey, D. Zwillinger, and S. Technica,
\emph{Table of integrals, series, and products,} 7th ed. New York, NY, USA:
Academic, 2007.

\bibitem{P.J.Davis}
P. J. Davis and P. Rabinowitz, \emph{Methods of numerical integration}, Academic press, 1984



\bibitem{Book:SG2013}
S. N. Chiu, D. Stoyan,W. S. Kendall, and J. Mecke, \emph{Stochastic geometry
and its applications,} 3rd ed. Hoboken, NJ, USA: Wiley, 2013.




\end{thebibliography}
\end{document}